\documentclass[sigconf]{acmart}
\pdfoutput=1  

\AtBeginDocument{%
  }

\usepackage{tikz}
\usepackage{amsmath}

\usepackage{comment}
\usepackage{booktabs}
\usepackage{xspace}
\usepackage{graphicx}
\usepackage{subfigure}
\usepackage{autobreak}
\usepackage{makecell}
\usepackage{diagbox}
\usepackage{listings}
\usepackage{graphics}
\usepackage{enumitem}
\usepackage{epsfig}
\usepackage{color}
\usepackage{ifpdf}
\usepackage{mathrsfs}
\usepackage{multirow}
\usepackage{xcolor}
\usepackage{url}
\usepackage{color,xcolor}
\usepackage{array}
\usepackage{multirow}
\usepackage{bm}
\usepackage{subfigure}
\usepackage{tabularx}
\usepackage{wasysym}
\usepackage{hyperref}
\usepackage{colortbl}
\usepackage{tabularray}

\usepackage{tcolorbox}
\usepackage[ruled]{algorithm2e}

\renewcommand\footnotetextcopyrightpermission[1]{}  

\newcommand{\sysname}{\textsc{VPVet}\xspace}
\newcommand{\datasetname}{\textsc{VRPP}\xspace}

\newenvironment{packeditemize}{
\begin{list}{$\bullet$}{
\setlength{\labelwidth}{8pt}
\setlength{\itemsep}{2pt}
\setlength{\leftmargin}{\labelwidth}
\addtolength{\leftmargin}{\labelsep}
\setlength{\parindent}{0pt}
\setlength{\listparindent}{\parindent}
\setlength{\parsep}{0pt}
\setlength{\topsep}{3pt}}}{\end{list}}

\newcommand{\bheading}[1]{{\vspace{0pt}\noindent{\textbf{#1}}}}

\begin{document}
\title{\sysname: Vetting Privacy Policies of Virtual Reality Apps}

\author{Yuxia Zhan}
\email{dabeidouretriever@sjtu.edu.cn}
\affiliation{%
  \institution{Shanghai Jiao Tong University}
 \city{Shanghai}
  \country{China}
}

\author{Yan Meng}
\authornote{Yan Meng is the corresponding author.}
\email{yan\_meng@sjtu.edu.cn}
\affiliation{%
  \institution{Shanghai Jiao Tong University}
 \city{Shanghai}
  \country{China}
}

\author{Lu Zhou}
\email{zhoulu@xidian.edu.cn}
\affiliation{%
  \institution{Xidian University}
 \city{Xi'an, Shaanxi}
  \country{China}
}

\author{Yichang Xiong}
\email{yxiong2@gmu.edu}
\affiliation{%
  \institution{George Mason University}
 \city{Fairfax}
 \state{Virginia}
  \country{USA}
}

\author{Xiaokuan Zhang}
\email{xiaokuan@gmu.edu}
\affiliation{%
  \institution{George Mason University}
 \city{Fairfax}
 \state{Virginia}
  \country{USA}
}

\author{Lichuan Ma}
\email{lcma@xidian.edu.cn}
\affiliation{%
  \institution{Xidian University}
 \city{Xi'an, Shaanxi}
  \country{China}
}

\author{Guoxing Chen}
\email{guoxingchen@sjtu.edu.cn}
\affiliation{%
  \institution{Shanghai Jiao Tong University}
 \city{Shanghai}
  \country{China}
}

\author{Qingqi Pei}
\email{qqpei@mail.xidian.edu.cn}
\affiliation{%
  \institution{Xidian University}
 \city{Xi'an, Shaanxi}
  \country{China}
}

\author{Haojin Zhu}
\email{zhu-hj@sjtu.edu.cn}
\affiliation{%
  \institution{Shanghai Jiao Tong University}
 \city{Shanghai}
  \country{China}
}

\renewcommand{\shortauthors}{Zhan et al.}

\begin{abstract}
Virtual reality (VR) apps can harvest a wider range of user data than web/mobile apps running on personal computers or smartphones. Existing law and privacy regulations emphasize that VR developers should inform users of what data are \textbf{c}ollected/\textbf{u}sed/\textbf{s}hared (CUS) through privacy policies. However, privacy policies in the VR ecosystem are still in their early stages, and many developers fail to write appropriate privacy policies that comply with regulations and meet user expectations. In this paper, we propose \sysname to automatically vet privacy policy compliance issues for VR apps. 
\sysname first analyzes the availability and completeness of a VR privacy policy and then refines its analysis based on three key criteria: granularity, minimization, and consistency of CUS statements. 
Our study establishes the first and currently largest VR privacy policy dataset named \datasetname, consisting of privacy policies of 11,923 different VR apps from 10 mainstream platforms. Our vetting results reveal severe privacy issues within the VR ecosystem, including the limited availability and poor quality of privacy policies, along with their coarse granularity, lack of adaptation to VR traits and the inconsistency between CUS statements in privacy policies and their actual behaviors.
We open-source \sysname system along with our findings at repository \url{https://github.com/kalamoo/PPAudit}, aiming to raise awareness within the VR community and pave the way for further research in this field.

\end{abstract}

\maketitle

\section{Introduction}
\label{sec-intro}

Enhanced by multi-modal sensors and binocular visual rendering techniques, virtual reality (VR) offers an unparalleled immersive experience for users, establishing itself as the fundamental technology for the meta-verse.
Major tech companies are heavily investing in VR headsets, with Meta releasing its most advanced Meta Quest Pro \cite{quest2_sold} and Apple developing their latest Apple Vision Pro \cite{apple_vision_pro}. 
As VR hardware continues to advance, there is a significant surge in the growth of VR applications (apps). Beyond its popularity in the gaming industry, VR apps are widely designed and deployed in various scenarios, including art~\cite{venice}, education~\cite{med-training}, healthcare~\cite{vr-health}, tourism~\cite{vr-tour}, military training~\cite{varjo-contract}, real estate~\cite{vr-real-estate}, retail~\cite{vr-shop}, sports~\cite{vr-nba}, and virtual meetings~\cite{vr-conference}. In August 2023, a VR company called Varjo closed a multi-million dollar deal to supply headsets for the US Army~\cite{varjo-contract}, showcasing the significance and promising future of VR technology.

Despite the widespread popularity of VR, researchers have found that the data collection/usage/sharing (CUS) process in VR apps faces an increasing risk of privacy leakage \cite{bailenson2018protecting, de2019security, bye2019ethical, nair2022exploring, jia2017ethical, happa2021privacy, buck2021privacy}. 
These privacy risks are more apparent in VR apps because, during the transition from deploying apps on desktop and traditional mobile devices (e.g., smartphones) to VR, the number and types of sensors and input/output (I/O) devices undergo significant increases, resulting in vast amounts of users' personal information being collected~\cite{adams2018ethics}.
For instance, the measurements from cameras or infrared trackers of VR headsets and hand controllers can reveal users' biometric information (e.g., height and body shape), while data from eye-tracking sensors can expose users' opinions toward virtual content (e.g., user interest in advertising content~\cite{meta-ad}).
Additionally, recent research indicates that users can be de-anonymized with over 90\% accuracy among a pool of 50,000+ individuals based on just 100 seconds of motion data in VR games~\cite{nair2023unique}.

Privacy concerns related to VR apps are not only drawing attention from academia but are also emphasized with the promulgation of privacy laws and regulations such as Personal Information Protection Law (PIPL) in China~\cite{PIPL}, General Data Protection Regulation (GDPR) in European Union~\cite{GDPR}, and California Consumer Privacy Act (CCPA) in United States~\cite{CCPA}. 
All of them stipulate that users have the \textit{right to know} about the data usage process. Specifically, \textit{privacy policies} serve as an important interface that allows users to understand how their personal data are collected, stored, and processed by any specific app in a \textit{transparent} manner.
Ideally, a well-written privacy policy can let users make informed decisions \textit{before} using these apps based on whether their personal information is being handled appropriately.

\noindent \textbf{Motivations.} 
Unfortunately, prior studies indicate that 91\% of users typically agree to privacy policies by simply clicking the checkbox, while skipping reading their contents~\cite{mcdonald2008cost,no-one-read-pp}. What's even worse, in the current VR ecosystem, our preliminary analysis (Section \ref{sec-motivation}) reveals that for a significant proportion of VR apps, their privacy policies neither meet the legal requirements nor satisfy user expectations, especially in the following three aspects. (1) \textit{Poor accessibility}: Some VR apps' developers and the platforms where these app are published do not display the corresponding privacy policy.
(2) \textit{Lacking VR-specific content}: even if the privacy policy is accessible, it lacks details on how user data is handled in VR scenarios. 
(3) \textit{Vagueness and misrepresentation}: the privacy policy either uses coarse descriptions or excessively claims to collect users' data, which deviates from apps' actual practices. The above observations motivate us to vet the compliance of privacy policies in the whole VR ecosystem.

\noindent \textbf{Challenges.}
However, vetting privacy policies in the VR domain must tackle the following research challenges.
\begin{packeditemize}
    \item \textbf{There is currently no unified criteria for vetting privacy policies, due to the decentralized and heterogeneous nature of VR platforms.}
    For a given type of VR device (e.g., Meta Quest 2), VR apps can be published on either its native platform (e.g., Meta) or compatible third-party platforms (e.g., SideQuest). However, the diverse regulatory requirements across heterogeneous app platforms (e.g., Meta requires a privacy policy for published apps while SideQuest does not) lead to varying quality in privacy policies, further complicating the setting of appropriate vetting criteria.
    
    \item \textbf{Existing vetting tools experience significant performance drops in VR domain.} 
    Natural language processing (NLP) driven privacy tools (e.g., PolicyLint~\cite{policylint}, PolicyGraph~\cite{poligraph}) leverage named entity recognition (NER) and terminologization to process privacy policy sentences, but their performance is fragile when facing domain changes (e.g., shifting from mobile phones to VR).
    The emergence of many new data objects in VR (e.g., avatar, eye tracking) has negatively impacted the performance of these tools.
    For example, the NER of the latest privacy policy analysis tool, PoliGraph \cite{poligraph}, only achieves an 87\% recall rate for VR-domain privacy policies sentences, a stark contrast to its 95\% recall rate on general-purpose privacy policies sentences. Current efforts to address domain changes either rely on purely manual checks (e.g., in smart home domain \cite{manandhar2022smart}) or lack comprehensive coverage of VR-specific terms (e.g., OVRSeen \cite{ovrseen} only covers 100 privacy policies in VR). 
\end{packeditemize}

\noindent \textbf{\sysname}.
To address the above challenges, we develop \sysname, a novel system for vetting privacy policies for VR apps. To establish the vetting criteria,  we survey the current VR platform market and select 10 mainstream platforms as our research focus. By crawling and collecting more than 11.9k apps' information, we construct the first and currently largest VR privacy policy dataset named \datasetname. {Then}, based on the case studies from five typical VR apps' privacy policies and three representative privacy laws, we summarize five criteria for vetting VR privacy policies. Specifically, these criteria includes \textit{availability} and structural \textit{completeness} of a privacy policy, as well as their CUS statements' \textit{granularity}, \textit{minimization} requirements and \textit{consistency} with actual behaviors.

To enhance domain-shift vetting performance, \sysname first automatically synthesizes a VR-domain policy sentences dataset. This includes 1.3k CUS sentences embedded with 267 unique VR domain phrases and an additional 14k non-CUS sentences. 
Utilizing this synthetic dataset, \sysname fine-tunes PrivBERT \cite{privbert-2021}, a privacy policy language model pre-trained on $\sim$1 millions general privacy policies, resulting in a recall rate increase from 87\% to 98.2\% for VR-domain CUS sentences compared to the latest tool PoliGraph \cite{poligraph}. Additionally, \sysname introduces a semantic similarity-based clustering method that expands the terminologization coverage in the VR field by adding 84 more data object terms and covering an extra 5.8k phrases compared to OVRSeen \cite{ovrseen}. Based on the larger number of data object phrases and more complicated terminologization, \sysname defines novel metrics (i.e., the lower bound and upper bound of privacy policy's claimed CUS) to assess the granularity of VR privacy policies and fairly analyze their minimization requirements and consistency with actual behaviors.

We leverage \sysname to measure privacy policies in \datasetname and obtain following main findings (details are presented in Section \ref{sec-finding}).  
(1) \textit{Inadequate availability and missing components:} platforms like Viveport and SideQuest have an availability rate even less than 1\%. Furthermore, 65.3\% privacy policies lack essential components, especially in relation to children's privacy statements. (2) \textit{Coarse-grained CUS sentences:} 14.7\% data objects (including sensitive data like health information) are not well specified. The disclosure of third-party collectors is even worse, with 93.5\% of them not specified. (3) \textit{Tendency of overbroad collecting information:} 85.1\% VR apps present overbroad collections (similar to trends observed in Android apps \cite{zhou2023policycomp}). In particular, big companies like Qantas and Emirates assign legacy privacy policies to their VR experience apps without any specification of VR traits. (4) \textit{Discrepancy between policy and code practices:} 78.0\% of apps we tested in \datasetname show inconsistency between privacy policies and their actual code behaviors. 
What exacerbates this issue is that platforms like Meta have relaxed the inspection requirements for consistency, thus magnifying privacy breaches.

\bheading{Contributions.} We make the following contributions:

\begin{packeditemize}
   \item We design \sysname to analyze privacy policies in the VR ecosystem, which covers the vetting criteria of availability, and structural completeness of the privacy policy, as well as granularity, minimization, and consistency of the CUS statements. Especially, the methodologies proposed by \sysname to handle the domain shifts in VR can be easily deployed for vetting privacy policies in other domains.
   
   \item We construct a large-scale dataset named \datasetname, containing 11,923 distinct VR apps' meta-info from 10 mainstream VR platforms, as well as 3,521 valid privacy policies and 1,096 VR apps' package files. This dataset will be available to the public to facilitate further research.
    
    \item Using \sysname, we conduct the first large-scale measurement of privacy policies in \datasetname. Our findings reveal significant mishandling and disregard for privacy in the current VR ecosystem, as well as the potential reasons behind these phenomena.
    
\end{packeditemize}

\bheading{Ethical consideration.} Our analysis only relies on the meta-info and privacy policies of VR apps, which are publicly available (i.e., from VR platforms and homepages of VR apps). The dataset collection procedure are under the approval of the institutional review board (IRB) of our institutions. 
All discovered privacy issues are reported to the corresponding VR app platforms. We open-source \sysname and the findings to the public on \url{https://github.com/kalamoo/PPAudit} to help the VR community better assess privacy policies. 
\section{Background}
\label{sec-background}

\begin{figure}[t]
    \centering
    \includegraphics[width=0.95\linewidth]{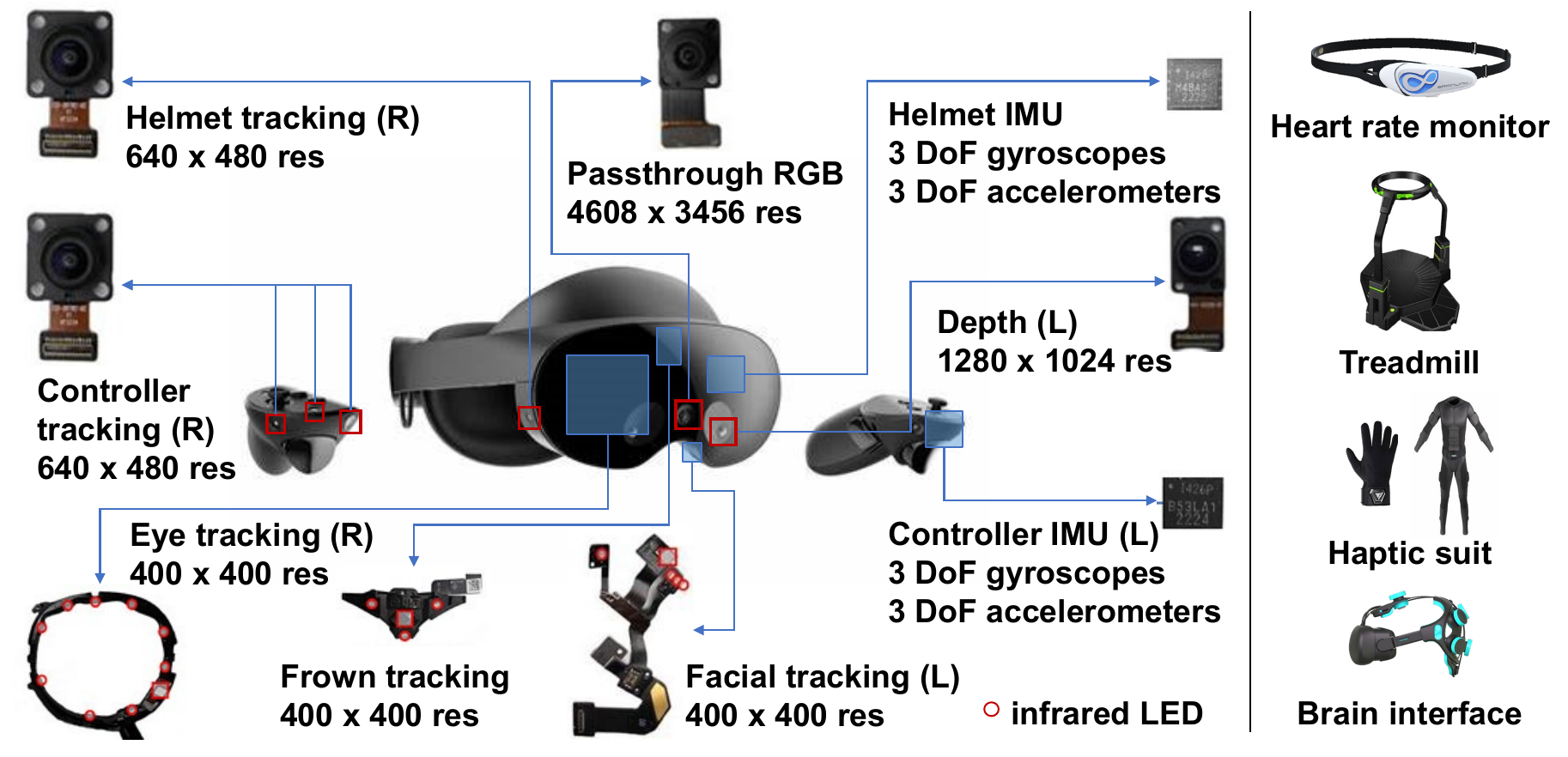}
    \caption{Sensors embedded in Meta Quest Pro (Left) and accessories emerging in the VR consumer market (Right).}
    \label{fig:vr-background}
\end{figure}

\begin{table*}[t]
    \centering
    \footnotesize
    \caption{Explanation and examples of components in the privacy policy. Examples are extracted from the privacy policy of VRChat, the most popular social VR app attracting more than 19 million active users \cite{vrchat_pp, vrchat_sold}.}
    \begin{tabular}{m{2.3cm} m{4.1cm} m{9.2cm}}
    \toprule
    \textbf{Component}          & Explanation                                                    & Example                                                                                                 \\ \midrule
    \textbf{Data CUS}    & data be collected/used/shared                    & \textit{we \textbf{may collect} or you \textbf{may provide} Personal Information about you}                                    \\ 
    \textbf{Data Retention}     & for how long user data is stored                               & \textit{when determining \textbf{the specific retention period}, we consider various factors}                         \\ 
    \textbf{Data Security}      & how user data is protected                                     & \textit{we use technical safeguards to \textbf{improve the integrity and security of} Personal Information}           \\ 
    \textbf{User Choice}        & options available to users                 & \textit{you \textbf{may opt out from} receiving commercial email by sending your request to us by email} \\ 
    \textbf{User Rights}        & users access/edit/delete their info & \textit{you may submit \textbf{a verifiable request that we delete} Personal Information about you}                   \\ 
    \textbf{Policy Change}      & how users be informed about changes & \textit{\textbf{if we modify} this Policy, we will make it available through the Platform}                            \\ 
    \textbf{Specific Audiences} & pertain to specific groups       & \textit{we do not knowingly collect Personal Information \textbf{from children}} 
    \\ \bottomrule                                     
    \end{tabular}
    \label{tab:vrchat_pp}
\end{table*}

\subsection{VR Devices}
\label{subsec-vr}

Virtual reality is a cutting-edge technology that provides users with a fully immersive experience, attracting major tech companies to invest in developing their own brand of VR headsets. Take one of the most advanced standalone VR devices, Meta Quest Pro \cite{meta_pro}, as an example, it features 16 cameras, 29 infrared LEDs, 3 IMUs, and many other sensors \cite{meta-pro-hardware} which are shown in Figure \ref{fig:vr-background}. Currently, there is a growing trend of integrating physiological and environmental monitoring accessories into VR devices \cite{vr-sensors-cao-2019}. Moreover, new accessories such as treadmills, haptic gloves, and brain interfaces are emerging with compatible consumer products already being introduced to the market. 

According to their work mode, VR devices can be classified into four types: \textit{(1) Optical Lens VR} consists of two convex lenses and several pieces of cardboards. 
\textit{(2) Smartphone VR} works by connecting to a mobile phone and acts as display screens as well as I/O devices. \textit{(3) PCVR} works much like Smartphone VR, but they are connected to a PC. \textit{(4) Standalone VR} devices can work independently, allowing users to play in any area without potentially dangerous cables. Some Standalone VR devices are compatible with PCVR or Smartphone VR mode. 
More details can be found on Section \ref{subsec-system-data-collection}.

\subsection{Privacy Policy in VR}
\label{subsec-bgr-pp}

\bheading{Requirements and expectations about privacy policy.}
A privacy policy is a statement that outlines how developers collect, use, share, and manage their users' data. The growth of privacy laws in recent years necessitates the privacy policy to function both as a notifications and a legal agreement between users and developers.
According to \cite{gupta2022creation}, 182 jurisdictions have enacted over 1,043 specialized privacy laws, some well-known ones including PIPL in China \cite{PIPL}, GDPR in Europe \cite{GDPR}, CCPA \cite{CCPA} and California Privacy Rights Act (CPRA) \cite{CPRA} in California, Personal Information Protection and Electronic Documents Act (PIPEDA) in Canada \cite{pipeda}, General Data Protection Law (LGPD) in Brazil \cite{lgpd} and Act on the Protection of Personal Information (APPI) in Japan \cite{appi}. Developers are responsible for complying with corresponding regulations by providing a publicly available privacy policy to their users and properly informing users of data collection as well as their privacy rights. A widely adopted category method \cite{wilson-2016-creation, pp-parts-ndss17, pp-parts-sec18, pp-parts-tse21, pp-parts-www18, pp-parts-www22} summarizes 7 components that a complete privacy policy is expected to include (See Table \ref{tab:vrchat_pp}). Note that we have merged the 1st and 3rd party data collection categories into \textit{Data CUS} while excluding \textit{Do Not Track (DNT)} and \textit{Others} categories.

\bheading{Early stage of privacy policy in the VR ecosystem.} 
It is reported that 91\% users agree to the privacy policy by clicking on the checkbox without necessarily reading it \cite{mcdonald2008cost, no-one-read-pp}. This phenomenon will have a more negative impact on VR scenarios: firstly, VR apps harvest more information about the user than existing mobile apps; secondly, some mainstream VR platforms pay little attention to vetting privacy policies and several platforms even do not require app developers to provide privacy policies when publishing apps. \textbf{Thus, it is desirable to design an effective and holistic tool to audit the privacy policies in the whole VR ecosystem.}

\section{Motivation and Vetting Criteria}\label{sec-motivation} 

The goal of this study is vetting the privacy policies in VR ecosystem. To justify the rationale behind our vetting method, we begin by presenting a hypothetical yet realistic scenario that users may encounter given current VR technologies\footnote{This story originated from 2018 XR Privacy Summit~\cite{Stanford-vr-summit}.}. Following this, we will summarize five vetting criteria for \sysname based on analyses of typical VR app privacy policy examples.

\begin{tcolorbox}[notitle, boxrule=0pt,left=0.2cm, right=0.2cm, top=0.1cm, bottom=0.1cm]
\textbf{Motivation scenario:} \textit{Riley participated in a virtual reality maze and solved puzzles by physically walking around her living room. She was able to interact with her friends using gestures such as nodding, high-fiving, and making eye contact. Unbeknownst to Riley, her 20-minute VR game session captured 2 million data points of her body movements, which were sold to an insurance company. As a result, the company denied Riley's life insurance policy due to her behavioral patterns resembling early dementia. Her sister was also rejected for insurance policies as dementia tends to run in the family...}
\end{tcolorbox} 

\begin{figure*}[t]
    \centering
    \includegraphics[width=0.98\linewidth]{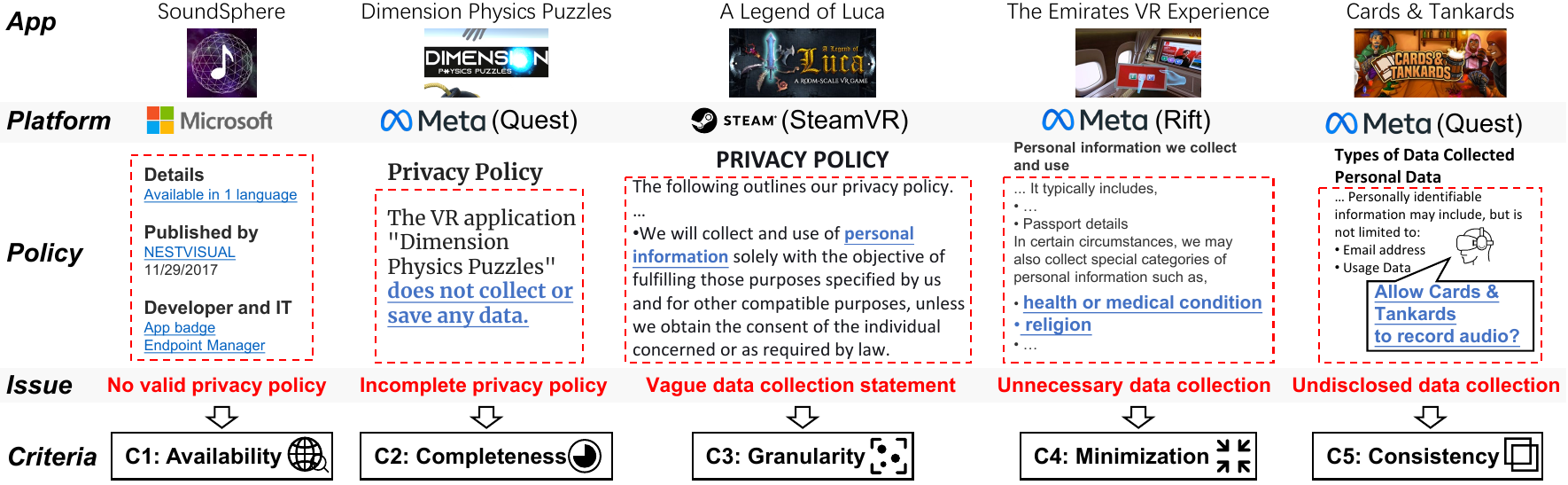}
    \caption{Motivation of \sysname: the current situation of VR app privacy policies and corresponding vetting criteria.}
    \label{fig:motivation}
\end{figure*}

Currently, there are no standards or regulations governing how VR data should be collected/used/shared. If Riley encounters the aforementioned situation where her data has been mishandled, she may first choose to read through the VR app's privacy policy for insights into what happened. However, Riley may face one of the following situations as shown in Figure~\ref{fig:motivation}, which inspired us to propose criteria for vetting VR apps.

Situation 1: She was unable to locate the privacy policy on either the app's homepage or the VR platform, like the case of the app \texttt{SoundSphere} published in the platform MicrosoftMR lacking a publicly available privacy policy.

As such, our first vetting criterion is \textbf{whether the VR app provides an easily accessible privacy policy for users (C1: Availability).}

Situation 2: She found the privacy policy, but it provides insufficient context for her to understand her privacy rights, like the case of the app \texttt{Dimension Physics Puzzles} published in Meta Quest. 

Therefore, our second vetting criterion is \textbf{whether the VR app's privacy policy contains the necessary structural components (C2: Completeness).}

Situation 3: She discovered that the privacy policy contains unclear statements regarding data collection (e.g., only stating personal information without further refinement in the privacy policy of \texttt{A Legend of Luca}, the best HTC Vive Games of 2016), raising her concerns about the specific type of personal information that will be collected or shared. 

Therefore, our third vetting criterion is \textbf{whether the VR privacy policy provides detailed and specific statements about data collection in a clear and precise manner (C3: Granularity).}

Situation 4: She found that the privacy policy claims to collect various types of data (e.g., medical conditions and religious info claimed in the privacy policy of \texttt{The Emirates VR Experience} published in the platform Rift) that do not seem necessary for the app's functionalities (i.e., VR experiences of a flight).

Therefore, our fourth vetting criterion is \textbf{whether the VR privacy policy only claims to collect the minimum amount of data required to support its functionalities (C4: Minimization).}

Situation 5: She remembered granting several sensitive permissions to the VR app (e.g., microphone permission in \texttt{Cards \& Tankards}, the most popular VR card game), but discovered that the privacy policy does not mention anything about it. 

Therefore, our final vetting criterion is \textbf{whether the VR privacy policy aligns with its actual behaviors (C5: Consistency).}

To summarize, there exists a significant disparity between the VR device's capacity to collect/use/share users' data and the insufficient transparency in informing users about it. With this motivation, we propose \sysname, the first automatic privacy policy vetting system for VR apps, to assess privacy policies based on the five aforementioned criteria.

\noindent \textbf{Justification between vetting criteria and legal requirements.} 
Some legal policies related to our vetting criteria are listed in Table \ref{tab:privacy_law} in Appendix~\ref{apdx-law}. 
It should be noted that failing to satisfy certain criteria does not necessarily mean that a privacy policy violates the corresponding legal articles listed in this table. Legal judgment is complex and requires consideration of several factors, making it beyond the scope of \sysname to provide legal advice or conclusions. The 5 vetting criteria and privacy reports generated by \sysname are intended only as recommendations and references to assist users in making informed choices and judgments.

\section{\sysname System}
\label{sec-system}

\begin{figure}[t]
    \centering
    \includegraphics[width=0.95\linewidth]{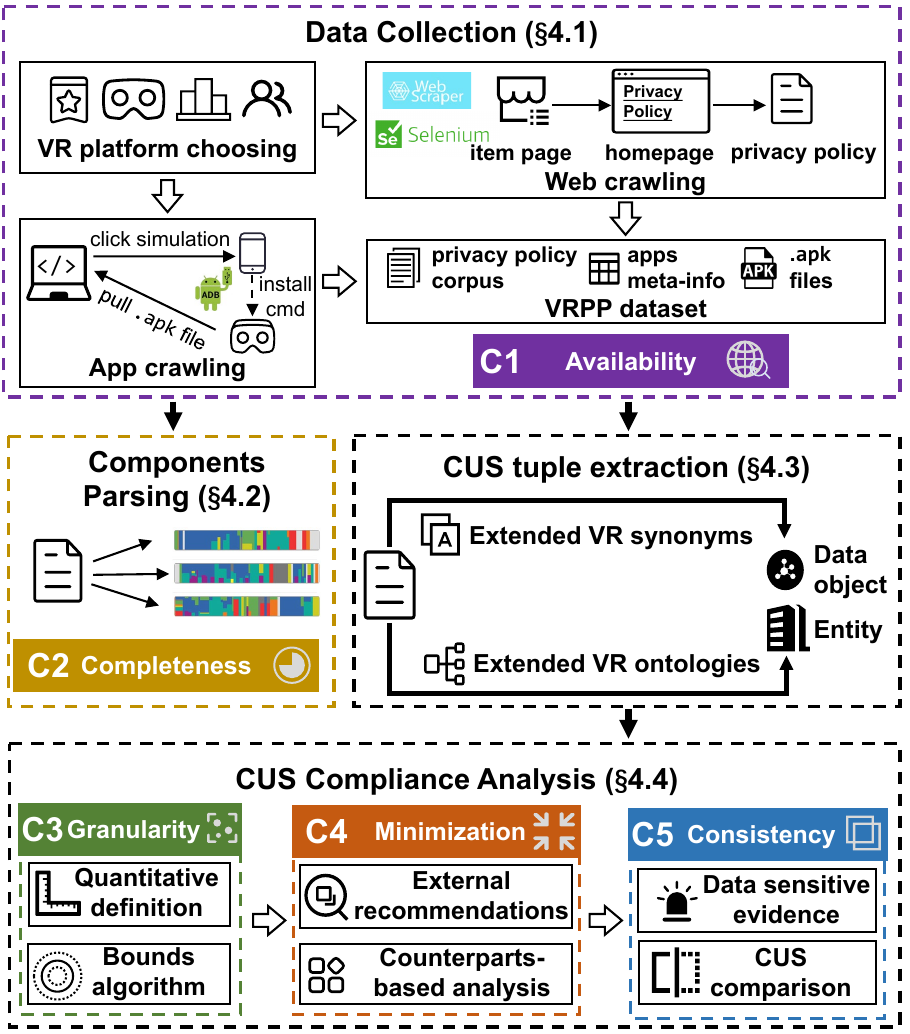}
    \caption{System overview of \sysname.}
    \label{fig:system}
\end{figure}

The workflow of \sysname is shown in Figure \ref{fig:system}. We first provide a brief introduction to our data collection process in the VR app ecosystem and the availability vetting process in Section \ref{subsec-system-data-collection}. Then, we demonstrate how to vet structural completeness in Section \ref{subsec-system-components}, followed by how to overcome the domain-shift challenges and accurately extract CUS tuples in Section \ref{subsec-system-cus}. Finally, in Section \ref{subsec-system-compliance}, we show how \sysname defines the granularity metric for privacy policies in VR and vets the minimization and consistency based on granularity analysis.

\subsection{Data Collection (Vetting C1)}
\label{subsec-system-data-collection}

To vet privacy policies, \sysname first needs to collect a large-scale dataset that can reveal the true situation in the current VR app ecosystem. We select suitable VR platforms and then use web crawling techniques to retrieve their information and privacy policies. Additionally, to conduct the consistency analysis, we select a subset of these platforms (i.e., Standalone VR) and crawl their package files.
The results of this module form the VRPP dataset, which is utilized in subsequent modules of \sysname. 

\begin{figure}[t]
    \centering
    \includegraphics[width=0.99\linewidth]{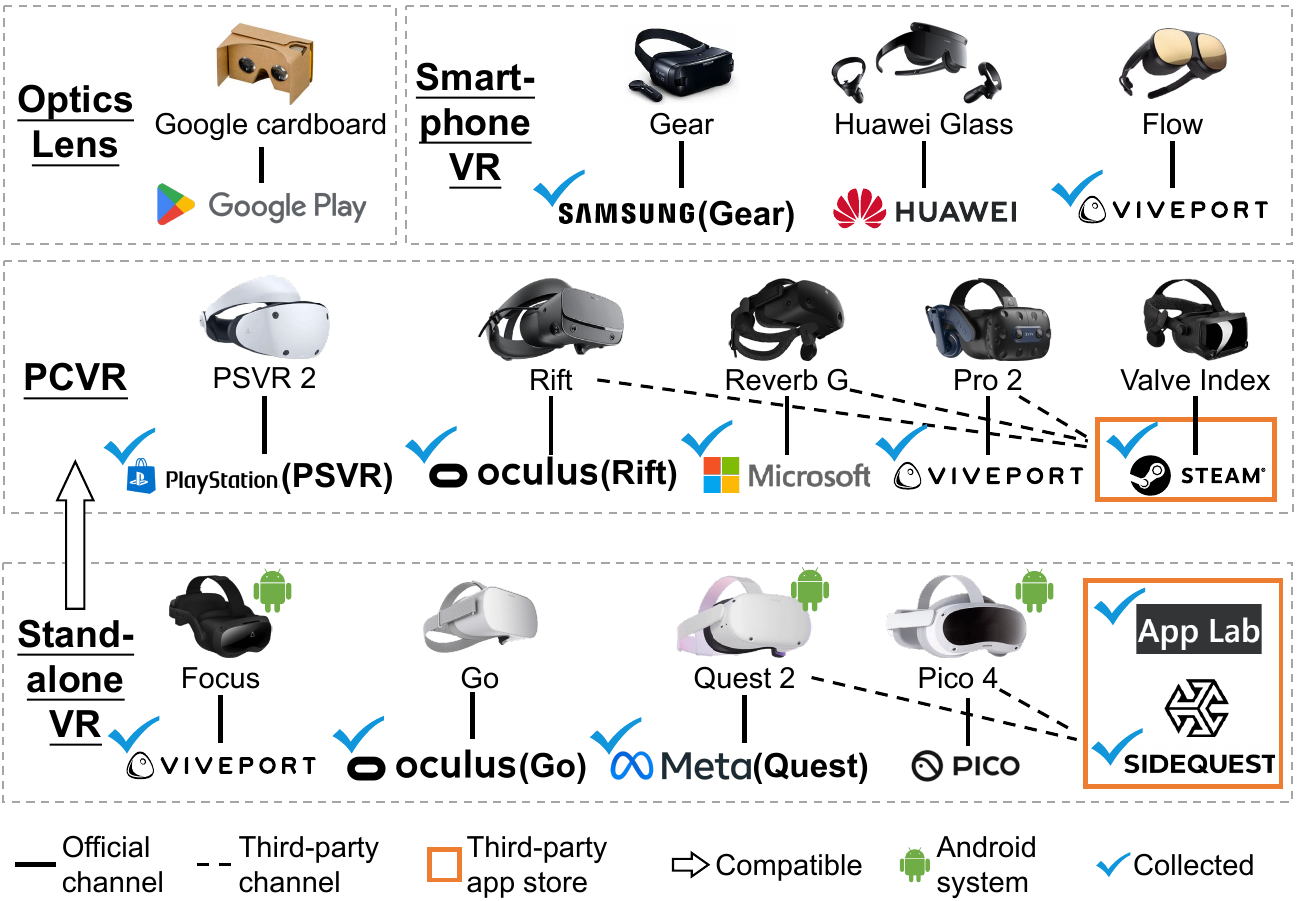}
    \caption{Mainstream VR devices and platforms.}
    \label{fig:vr-platform}
\end{figure}

\bheading{VR platform selection.} As depicted in Figure \ref{fig:vr-platform}, VR platforms exhibit heterogeneous trends. Typically, each VR device is associated with a native VR content platform, alongside several third-party platforms available. For example, a Quest 2 owner can not only download applications from Meta Quest store, but also from third-party stores such as Sidequest or App Lab. Hence we start by searching for top-selling VR devices and include their native VR platforms as candidates. Next, we aggregate these candidates with compatible third-party platforms. 
Finally, we exclude Optics Lens VR and its corresponding platform, e.g., Google Play for the Google cardboard; 
we also exclude platforms that do not have public websites (e.g., Pico and Huawei VR, whose content is only accessible to VR device owners). After de-duplication, this process yields a total of 10 popular VR content platforms. 

\bheading{Extracting VR app meta-info and privacy policies.} We utilize WebScraper \cite{web_scraper} and Selenium \cite{selenium} to crawl the meta-info (e.g., app name, app description, etc.) and privacy policy link of each app listed on its item page (i.e., the VR app's info-page in certain platform).
In cases where platforms do not provide direct links to privacy policies, we follow the homepage link (if available) of the app and extract clickable link objects whose text contains \textit{privacy} as their privacy policy link. Finally, we download the HTML document for each privacy policy link using \texttt{urllib} \cite{urllib} and convert them into plaintexts using  \texttt{HtmlToPlaintext}~\cite{HtmlToPlaintext}.

\begin{figure*}[t]
    \centering
    \includegraphics[width=0.99\linewidth]{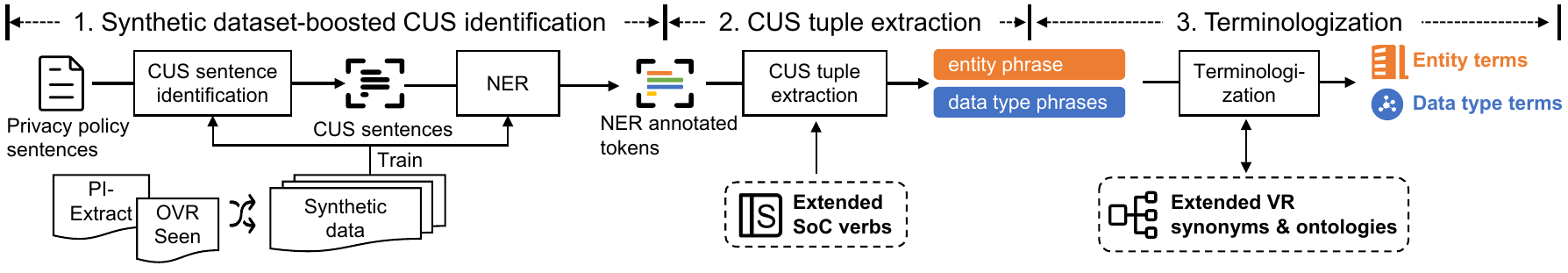}
    \caption{CUS tuple extraction pipeline.}
    \label{fig:csp-extraction}
\end{figure*}

\bheading{Downloading standalone VR apps.} 
Since only Standalone VR devices have their VR-specific apps while PCVR devices run desktop applications, we restrict the scope of app downloads to three Standalone VR platforms: the Official Quest store, its third-party app platform SideQuest, and the semi-official App Lab. 
For SideQuest apps, we use \texttt{GetSidequestURL} \cite{GetSidequestURL} to download the app's package file through its source URL. For Meta Quest and App Lab platforms, since the app can only be downloaded through the official Quest app store on the VR headset, we automate this downloading process by simultaneously controlling a rooted Android mobile phone (with Meta Quest app installed and logging into our account) and the paired Quest device with a script running on PC. 

In order to responsibly collect data and avoid impacting the VR platform server, we select a relatively slow speed and frequency of our automatic crawling method. The interval between each click for the web browser is longer than 10 seconds; the interval between each download command for the app's package file is longer than 2 minutes; and the execution time of the command depends on the size of the app. Throughout our data collection procedure, we did not receive any complaints or warnings from these platforms.

\noindent \textbf{Vetting availability (C1):} we consider a privacy policy to be easily accessible and therefore \textit{available} if it is provided either (1) on the item page of the VR platform, or (2) on the homepage of the VR app within two-hop. For a specific VR app platform, we define its availability as the ratio of published apps that have available privacy policies.
The detailed vetting result are described in \textbf{\textsc{Findings} \large 1} and \textbf{\large 2}  of Section~\ref{subsec-primary-finding}.

\subsection{Components Parsing (Vetting C2)}
\label{subsec-system-components}

To vet structural completeness of a privacy policy, \sysname needs to parse the privacy policy into different components. Specifically, \sysname labels each sentence in the privacy policy according to its component category (as listed in Table~\ref{tab:vrchat_pp}) in the privacy policy. Due to the complexity of legal document text, a sentence may contain more than one type of statement. For instance, in the sentence \textit{``we may collect your IP address, if you do not want us to do so, you can opt-out''}, both \textit{Data CUS} and \textit{User Choice} components are presented. Therefore, we model this classification problem as a multi-label task.  

For this task, we deploy an end-to-end component parser consisting of a privacy policy language model called PrivBERT \cite{privbert-2021} (see Section \ref{subsubsub-cus-id-and-ner} for more details) and a subsequent multi-label classification model. 
We train our parser on the OPP-115 dataset \cite{wilson-2016-creation} using a 6:2:2 split for training, validation, and testing, employing the same hyper-parameters as PrivBERT. The components parser achieves an F1 score of 81.3\% on the OPP-115 test set. To evaluate its performance in our \datasetname dataset, we randomly select 200 sentences and manually evaluate classification results. The parser obtained a macro recall rate of 91.0\%, demonstrating its efficiency.
Finally, we deploy this parser to analyze all privacy policies in the \datasetname dataset.

\noindent \textbf{Vetting completeness (C2):} A privacy policy is considered as \textit{structurally complete} if it includes all 7 components (listed in Table~\ref{tab:vrchat_pp}), and \textbf{\textsc{Finding} \large 3} of Section~\ref{sec:result:completeness} shows the vetting result.

\subsection{VR Domain-adapted CUS Extraction}
\label{subsec-system-cus}

\label{subsec-cus}

To vet the compliance of CUS statements in the privacy policy, we first have to extract CUS tuples. A CUS tuple $\langle e, d\rangle$ comprises two components: an entity $e$ and a data type $d$, indicating that the entity $e$ has collected, used, or been shared with type $d$ of data from users. The pipeline of extracting CUS tuples is shown in Figure \ref{fig:csp-extraction},  consisting of 4 main steps, i.e., CUS sentence identification, Named entity recognition (NER), CUS tuple extraction and Terminologization. 

However, the domain changes from mobile apps' privacy policies to VR apps' privacy policies can negatively affect these steps' performance, especially the recall rate of final VR specific terminologized CUS tuples. For example, for the CUS sentence \textit{``We may collect your ip address''}, the CUS tuples within it can be extracted and terminologized by privacy policy analysis tools like OVRSeen \cite{ovrseen} (which is based on PolicyLint \cite{policylint}) and PoliGraph \cite{poligraph}. However, if we changes the CUS sentence to \textit{``We may collect your eye tracking data''}, the corresponding CUS tuples although can be extracted but cannot be successfully terminologized. This is because even the VR oriented privacy policy tool OVRSeen doesn't include or record \textit{eye tracking data} in its synonyms files and ontologies\footnote{The ontology is a directed tree, which serves to represent the subsumptive relationships between terms, where a link from term $A$ to term $B$ indicates that $A$ is a broader term (hypernym) that subsumes $B$. More details can be found on Section \ref{subsubsub-term}.}. If we further complicate the CUS sentence with semantic structure like \textit{``There may also be opportunities for you to grant permission for use of other of your eye tracking information''}, then the latest general purpose privacy policy tool PoliGraph cannot recognize the data type using their NER.

\textbf{Therefore, \sysname proposes a collective of techniques to enhance the VR-domain adaptability of this CUS tuple extraction pipeline.} These includes: (1) constructing a synthetic dataset to boost the performance of CUS sentence identification and NER, (2) extending the share and collect (SoC) verbs for better CUS tuple extraction and (3) proposing a semantic similarity-based clustering method to efficiently enlarge the coverage of existing VR synonyms and ontologies.

\subsubsection{Synthetic Dataset-boosted CUS Sentence Identification and NER Models.} \label{subsubsub-cus-id-and-ner} 
\sysname first identifies those sentences that claim to collect/use/share users' data (\textit{CUS sentences}) from the privacy policy text and then label data object and entity tokens within each CUS sentence. Specifically, \sysname utilizes a synthetic dataset combined with the OVRSeen \cite{ovrseen} synonyms file (with 267 VR domain phrases of 41 different types extracted from 100 VR privacy policies) and PI-Extract \cite{bui2021automated} dataset (with fine-grained manually annotated labels of data objects from 30 privacy policies) to train these two models. We retained sentences that do not contain any labels of data objects (\textit{non-CUS sentences}) and considered others as \textit{candidate CS sentences}. We then inserted VR domain phrases into the labeled position within these candidate CS sentences. In total, we obtained 14k non-CUS sentences and 1.3k CUS sentences embedded with VR domain terms. We train the models on the synthetic dataset using a 6:2:2 split for training, validation, and testing. As a result, the CUS sentence identification model achieves a F1 score of 82.0\%, while the NER model achieved a F1 score of 86.5\%. For comparison, the latest domain-shift NER for smart home privacy policies \cite{manandhar2022smart} achieves an F1 score of 75.75\% with the assistance of a manually curated dataset of 284 privacy policies on that domain.

We attribute this performance improvement to the following two reasons. First is the \textbf{privacy policy language model} we use, i.e., PrivBERT that is pre-trained on millions of privacy policies which enables it to better capture features of privacy policies. Masked Language Modeling (MLM) task on \datasetname-Corpus results indicate that PrivBERT has a better perplexity (8.59) compared to a general-purpose language model like distilroberta-base (10.86) used in PoliGraph \cite{poligraph}. Second is the \textbf{high-quality synthetic dataset}, which leverages VR-domain knowledge and enables the model to better recognize VR-domain entities. Additionally, the CUS sentences it utilizes is constructed from real-world privacy policies which often have more complex syntactic structures than simple statements like \textit{``We will collect your <data>''}. This helps the model recognize unseen entities as long as they appear in the proper syntactic slot in these CUS sentences. To evaluate this, we selected 47 VR-domain data objects (e.g., \textit{body measure} and \textit{arm length}) and 42 general data objects (e.g., \textit{email address} and \textit{age}), and insert them into the CUS sentences to test the latest general-purpose NER in PoliGraph \cite{poligraph} and our model. The results demonstrate that PoliGraph-NER can detect an average of 95.0\% general data objects and 87.0\% VR-domain data objects, while our models can achieve a detection rate of 96.5\% for general cases and 98.2\% for VR-domain cases. Therefore, this method avoids manually annotating hundreds of VR domain privacy policies while achieving great performance in handling VR-specific phrases.

\begin{table}[t]
    \centering
    \footnotesize
    \caption{Extended SoC verbs used by \sysname.}
    \begin{tabularx}{\linewidth}{p{1.1cm} | X}
        \toprule
         \textbf{Type} & \textbf{Word} \\
         \midrule
         Sharing & \makecell[l]{disclose, distribute, exchange, give,
         provide, rent, report, \\
         sell, send, share, trade, transfer, transmit, \textbf{pass}, \textbf{express}, \\
         \textbf{supply}, \textbf{display},\textbf{deliver}, \textbf{release}, \textbf{publish}, \textbf{lease}, \textbf{download}, \\
         \textbf{reveal}, \textbf{tell}, \textbf{view}, \textbf{show}, \textbf{hold}, \textbf{swap}, \textbf{forward}} \\
         \hline
         Collection & \makecell[l]{access, check, collect, gather, know, obtain, receive, save, \\
         store, use, \textbf{get}, \textbf{perform}, \textbf{analyze}, \textbf{process}, \textbf{log}, \textbf{keep}, \textbf{add}, \\ 
         \textbf{record}, \textbf{combine}, \textbf{retain}, \textbf{recognize}, \textbf{track}, \textbf{remember}, \\ 
         \textbf{relate}, \textbf{create}, \textbf{ask}, \textbf{conduct}, \textbf{monitor}, \textbf{request}, \textbf{link}, \\ 
         \textbf{associate}, \textbf{solicit}, \textbf{read}, \textbf{preserve}, \textbf{contain}} \\
         \bottomrule
    \end{tabularx}
    \label{tab:soc_verbs}
\end{table}

\subsubsection{CUS Tuples Extraction with Extended SoC Verbs.} Given that a sentence may contain multiple data objects and entity tokens, we leverage the data and entity dependency (DED) tree from PolicyLint \cite{policylint} to establish a mapping between the data object and their corresponding entities based on their grammatical relationships within the context. We also expand the list of sharing or collecting (SoC) verbs from 23 words to 64 words (see Table \ref{tab:soc_verbs}) to improve the recall rate of the DED tree. These improvements have resulted in the discovery of 54.5\% (125,909/231,106) newly identified CUS tuples from \datasetname-Corpus.

\subsubsection{Terminologization.} \label{subsubsub-term} It is flexible for privacy policies to employ different phrasings when referring to the same data object or entity. For instance, phrases such as \textit{record of your voice instruction} and \textit{voice clip} both stand for data object term \textit{audio}. Terminologizing these semantically similar phrases (i.e., \textit{synonyms}) can help simplify subsequent analysis on CUS compliance. 
Existing works either uses manually curated lists of synonyms (PolicyLint \cite{policylint} and OVRSeen \cite{ovrseen}) or uses patterns-based method (PoliGraph \cite{poligraph}) to terminologize phrases. However, these methods either require massive human labor on checking every extracted phrases or will miss a significant number of phrases that do not comply with pre-defined patterns. For instance, when we utilize OVRSeen \cite{ovrseen} (PoliGraph \cite{poligraph}) to terminologize extracted CUS tuples from the \datasetname-Corpus, 7,069 (8,419) data object phrases, which covers 42,514 (49,254) CUS tuples, are reported as un-terminologized. Therefore, we propose an approach based on the insight that \textit{synonyms have similar semantics and will therefore be clustered in the embedding space}. \sysname first utilizes a BERT-based sentence embedding model to map all phrases to the semantic embedding space. During this phase, any phrases that are within a threshold distance (0.8 in our study, i.e., the median similarity of the OVRSeen synonyms file) are added to synonym lists. For remaining un-terminologized phrases, we iteratively spot new clusters in embedding space and determine whether and where they can be included in the VR data ontology. Additionally, we employ keyword matching method to handle entity phrases that lack semantic meaning (such as the names of apps, developers, or domains). More details be found on Appendix \ref{apdx-cus-pipeline}. As a result, we obtain an extended VR data (entity) ontology with 107 (117) nodes along with synonym lists containing 8,042 (1,663) distinct phrases (See Figure \ref{fig:onto} of Appendix~\ref{apdx-cus-pipeline} and summary of changes in Table \ref{tab:onto-sum}). 

\begin{table}
    \centering
    \caption{Comparison of OVRSeen and \sysname ontologies and synonyms.}
    \footnotesize
    \begin{tabular}{llll}
    \toprule
        Platform & OVRSeen & \sysname & \begin{tabular}[c]{@{}l@{}}New-in \\ \sysname\end{tabular} \\
    \midrule
        Data ontology & 63 & 107 & 84 \\
        \#Data synonyms & 2009 & 8042 & 5861 \\
        Entity Ontology & 64 & 117 & 60 \\
        \#Entity synonyms & 894 & 1663 & 969 \\
    \bottomrule
    \end{tabular}
    \label{tab:onto-sum}
\end{table}

\subsection{CUS Compliance Analysis} 
\label{subsec-system-compliance}

\subsubsection{Granularity Analysis (\textbf{Vetting C3})}

In some cases, privacy policies provide vague statements regarding data collection, rather than specifying the exact types of data that being collected, used or shared. For instance, compared to a CUS sentence saying it will \textit{``collect your health information''}, one stating it will \textit{``collect your health information such as your workout data''} provides users with a clearer understanding of what data is being collected. However, to the best of our knowledge, no existing metrics can quantitatively measure the granularity aspect of CUS statements in a privacy policy. To address this, \sysname introduces two metrics: \textit{CUS tuple granularity}, which measures the granularity of the entity and the data type term in a give CUS tuple; and \textit{privacy policy granularity}, which measures the granularity of a privacy policy by the set of data types they claim to collect. Their definitions are given below.

\begin{figure}[t]
    \centering
    \includegraphics[width=0.99\linewidth]{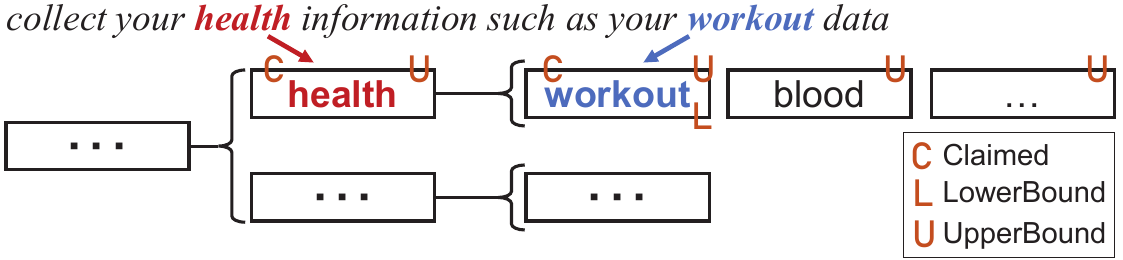}
    \caption{Granularity illustration. This is a snippet from the complete data ontology shown in Figure \ref{fig:onto}.}
    \label{fig:bounds-example}
\end{figure}

\bheading{CUS tuple granularity (CTG).} As illustrated in Figure~\ref{fig:bounds-example}, for every node $v$ (data object or entity) in the VR ontology $O$, the closer to the leaf node, the more fine-grained it is. Hence, we define its granularity as the longest distance from it to any leaf node: 
\begin{equation}
    \mathit{CTG}_O(v) \overset{\underset{\mathrm{def}}{}}{=} \max_{s \ \in \ S}{\left ( \max_{p\ \in \ \mathit{Paths}_O(v,s)}{(\mathit{Len}_O(p))} \right ),}
\end{equation}
where $S$ is the set of all leaves in $O$, $\mathit{Paths}_O(v, s)$ is the set of all simple paths from node $v$ to leaf $s$ in $O$, and $\mathit{Len}_O(p)$ is the length of path $p$ (defined as the number of nodes in this path) in $O$. The range of CTG of a node in VR data or entity ontology (See Figure \ref{fig:onto}) is from 1 (the leaf) to 5 (the root). The smaller the CTG, the more fine-grained the term.
For example, in Figure \ref{fig:bounds-example}, $CTG_O(\textit{workout}) = 1$ while $CTG_O(\textit{health}) = 2$.
Note that, for a given CUS tuple, since it contains both an entity $e$ and a data object $d$ , the calculated CTG is a 2-tuple $ \langle CTG_O(\textit{e}), CTG_O(\textit{d}) \rangle$.

\bheading{Privacy policy granularity (PPG).} 

We define this based on the following two insights. \textit{Firstly,} claimed CUS tuples in a privacy policy have two bounds: the lower bound (data types that are explicitly stated as being collected) and the upper bound (potential data types that may be collected). For example, in the CUS sentence \textit{``we may collect your health information such as your workout data''}, \textit{workout data} is explicitly claimed to be collected while the other types of data like \textit{blood sugar} will be potentially collected because they are the child of the \textit{health} node in the ontology. Figure \ref{fig:bounds-example} illustrates the lower and upper bounds of this example. See Algorithm \ref{alg:lower-bounds} and Algorithm \ref{alg:upper-bounds} in Appendix for more details on how to calculate them.
These bounds satisfy the following inequation: 
\begin{equation}
    \varnothing \sqsubseteq \mathit{LowerBound} \sqsubseteq \mathit{Claimed} \sqsubseteq \mathit{UpperBound} \sqsubseteq \mathbb{U}_T,
\end{equation}
where $\mathbb{U}_T$ is the set of all terms in the ontology. Secondly, if the gap between the lower and upper bounds is small, we consider an app's privacy policy to be fine-grained, as there are no ambiguous interpretations for these statements. Therefore, we define the PPG of the given privacy policy as
\begin{equation}
    \mathit{PPG}  = \mathit{UpperBound} -\mathit{LowerBound},
\end{equation}
The range of PPG is from 0 to the number of nodes in the data ontology (i.e., 107 in this study). The smaller the PPG, the more fine-grained the privacy policy.

\noindent \textbf{Vetting granularity (C3):} We regard the CUS tuple granularity as coarse-grained if the CTG is equal or larger than 2. As for the privacy policy granularity, considering PPG is a relative value, we will not set a threshold to determine whether an entire privacy policy is fine-grained or not. Instead, we'll provide PPG's distribution in \textbf{\textsc{Finding} \large 4} of Section \ref{subsec-deep-finding} and PPG's ranking percentiles in each app's privacy policy report. We also provide the vetting results regarding granularity aggregated by different VR platforms.

\subsubsection{Minimization Compliance Analysis (\textbf{Vetting C4})}
\label{subsec-c4}

Minimization compliance requires a privacy policy to only collect necessary data from users. However, there is no legal definition of \textit{minimization} specifically for various apps. To address this issue, we adopted the counterpart-based method described in \cite{zhou2023policycomp}, which compares CUS between a target app and similar apps (referred to as \textit{counterparts}). The insight behind this approach is that, if a group of apps provide similar functionalities, then they are expected to collect similar scope of data to support those functions. Under this assumption, any CUS data that is not collected by the majority of counterparts can be considered unnecessary, i.e., \textit{overbroad}. Considering that privacy policies have different granularity levels when referring to a data type, we perform such counterparts-based comparisons based on their $\mathit{LowerBound}$ of CUS statements. 

To find proper counterparts for each target app, we propose a multi-sources counterpart searching that considers (1) direct recommendations within a single platform (such as SteamVR), (2) cross-platform recommendations from professional recommend websites (like steampeek), (3) app genres, and finally determine the top-k counterparts based on their  (4) app descriptions.

\noindent \textbf{Vetting minimization (C4):} We consider a privacy policy to meets the minimization criterion if none of its claimed CUS tuples has overbroad data type. Vetting results regarding minimization aggregated by data types as well as aggregated by different VR platforms can be found in \textbf{\textsc{Finding} \large 5} of Section \ref{subsec-deep-finding}.

\subsubsection{Consistency Compliance Analysis (\textbf{Vetting C5})}
The app is expected to disclose all collected data objects in its privacy policy to ensure consistency compliance. As we do not have access to the server database of the app, we examine its behavior based on its code by de-compiling source code from their apk files.

We focus on data-sensitive evidence in the source code. This evidence includes sensitive permissions required by the app and data-sensitive functions/methods/APIs/URIs. We manually constructed a mapping file that illustrates the mapping relation between data objects and sensitive permissions and APIs on VR apps based on \cite{api-mapping2016}. We first updated this mapping to be compatible with Android 10 and Android 12, which most standalone VR devices are built upon. Third-party VR APIs such as Oculus, WaveVR, and Samsung are also included to cover more VR-domain data-sensitive behaviors. In total, we construct a mapping from 167 data-sensitive evidences to 28 data objects (15 of which are of VR traits). Finally, we compared the data-sensitive evidence in code with CUS tuples extracted from an app's privacy policy to examine their consistency. 

\noindent \textbf{Vetting consistency (C5):} We consider data objects that have evidence in the app's code to be vaguely claimed if they are in $\mathit{UpperBound} - \mathit{Claimed}$ set. 
If a data object is not even covered by the $\mathit{UpperBound}$, then we regard it as inconsistent. The results can be found in \textbf{\textsc{Finding} \large 6} in Section \ref{subsec-deep-finding}.
\section{Vetting Results}
\label{sec-finding}

\subsection{Dataset Description}
\label{subsec-dataset-des}

\begin{table}[t]
\caption{VRPP-Corpus description.}

\centering
\footnotesize
\label{tab:dataset-des}
\begin{tabular}{cccc}
\toprule
Platform & \# APP Info & \# PP & \begin{tabular}[c]{@{}c@{}}$\frac{\# PP}{\#  App Info}$\end{tabular} \\ \midrule
Sidequest & 2,274 & 192 & 0.084 \\
Viveport & 2,919 & 281 & 0.096 \\
PSVR & 580 & 91 & 0.157 \\
SteamVR & 6,748 & 1,185 & 0.176\\
Microsoft & 285 & 83 & 0.291 \\
Gear & 1,085 & 1,083 & 0.998 \\
Go & 1,118 & 1,116 & 0.998 \\
Rift & 1,368 & 1,366 & 0.999 \\
Quest & 387 & 387 & 1 \\
App Lab & 1,335 & 1,335 & 1 \\ \hline 
De-duplicated Summary & \textbf{11,923} & \textbf{3,521} & \textbf{0.295} \\ \bottomrule
\end{tabular}
\end{table}

\begin{table}[t]
\caption{VRPP-APK description.}
\centering
\footnotesize
\label{tab:vrpp-apk-des}
\begin{tabular}{ccccc}
\toprule
Platform & \# APP Info & \# PP & \# APK & \# APK w/ PP \\ \midrule
Sidequest & 1,319 & 83 & 691 & 46 \\
Quest & 116 & 108 & 41 & 36 \\
App Lab & 981 & 929 & 364 & 349 \\ \hline 
De-duplicated Summary & \textbf{2,416} & \textbf{928} & \textbf{1,096} & \textbf{286} \\ \bottomrule
\end{tabular}
\end{table}

Our dataset \datasetname consists of two part: \datasetname-Corpus and \datasetname-APK. \datasetname-Corpus is collected during September 2022 and October 2022, including meta-info of 17,299 VR apps from 10 mainstream VR platforms. Meta-info includes apps' common attributes, such as name, platform, item page link, homepage link (if any), privacy policy link, publisher details, genre description, and price. Considering that a VR app may be published on more than one platform, we aggregate (de-duplicate after merging) information for identical apps (by name), resulting in a total of 11,923 unique apps. After preprocessing the privacy policy links to extract plaintexts, we obtained 3,521 valid privacy policy texts for further analysis. The distribution of apps and privacy policies in different VR platforms is shown in Table \ref{tab:dataset-des}.

Since \datasetname-Corpus can only support vetting for criteria C1\textasciitilde C4, to vet consistency (C5), we additionally collect VR package files and construct \datasetname-APK, during September 2023. During dataset construction, we focus on free apps from Meta Quest, Sidequest, and App Lab platforms. This is the subset of the aforementioned dataset with an app file in \texttt{.apk} format. In total, we collected meta-info from 2,416 apps and obtained 928 valid privacy policies as well as 1,096 apk files (See \ref{tab:vrpp-apk-des}). Among them, there are in total 286 VR apps with both \texttt{.apk} file and valid privacy policy downloaded. It should be noted that we failed on approximately 40.0\% of the apk download links when trying to get the valid apk files on Sidequest platform. Moreover, there is a failure rate of 66.7\% for downloading apps' apk files from Quest headset for Quest and App Lab platforms. We will discuss this limitation on Section \ref{sec-limitation}.

\subsection{Vetting Results of Availability}
\label{subsec-primary-finding}

\textbf{\textsc{Finding} \large 1}: \textbf{Several VR platforms, including major ones like PSVR and Steam VR, due to inadequate regulations, have poor availability (less than 0.3) of privacy policies.}

As shown in Table \ref{tab:dataset-des}, for totally 11,923 VR apps we crawled on VR platforms, only a small portion (i.e., 29.5\%) of privacy policies were successfully found. Specifically, half of the mainstream platforms in the \datasetname-Corpus, named Viveport, Microsoft, PSVR, SteamVR, and Sidequest have low availability rates. On the contrary, App Lab, Quest, Go, Rift, and Gear - all under Meta's supervision - have high availability ratios with values close or equal to 1. Some real-world examples of VR apps fail to provide privacy policies can be found on Table~\ref{tab:C1-violate} of Appendix~\ref{sec:example:violate}.

\bheading{Potential reasons of Finding 1}. We investigated the reasons behind this phenomenon and discovered that whether developers provide a privacy policy for their VR app largely depends on the platform's requirements. Meta provides guidelines to developers \cite{meta-pp-requirement}, in which they require developers to \textit{``maintain a publicly available link to your privacy policy ... and ensure the link remains current and up to date''}. This explains why platforms like App Lab, Quest, Go, Rift that are regulated by Meta has availability rate large than 99.8\%. Similar guidelines can also be found on Microsoft and Viveport, however, they are not mandatory. Microsoft designates a \textit{Privacy Policy URL} as \textit{Required} in their checklist but notes that it is \textit{Sometimes not require} for publishing an app \cite{micro-guideline}. Viveport states that \textit{``If you (developer) have your own privacy policy, you may enter its URL.''} but does not make it compulsory \cite{vive-guideline}. For the Sidequest platform, which performs worst in the availability of the privacy policy, we attribute it to its third-party status. In their terms \cite{sidequest-terms}, they claim to be \textit{``a listing service only and have no responsibility for the content or accuracy, completeness, or lawfulness of the listings or the games''}. \\

\noindent \textbf{\textsc{Finding} \large 2}: \textbf{
Privacy policy reuse is a common issue for VR apps, with 54.5\% of vetted policies being shared among different apps, despite variations in their data collection practices.
}

\begin{figure}[t]
    \centering
    \includegraphics[width=\linewidth] {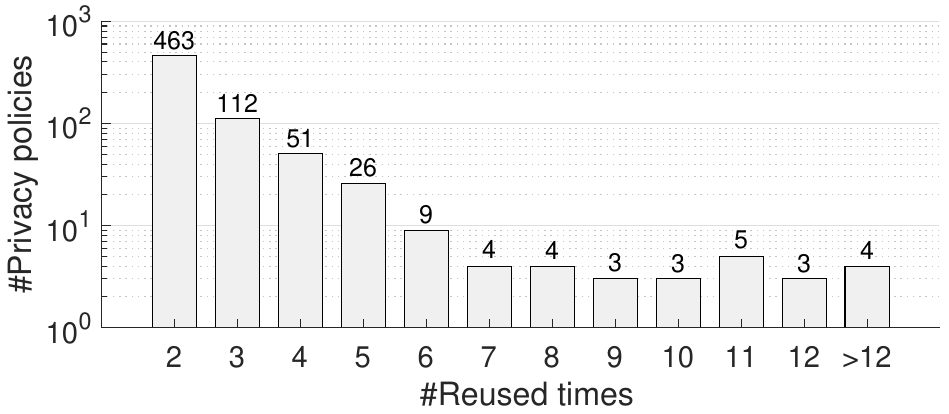}
    \caption{Reused privacy policies distribution.}
    \label{fig:finding_pp_reuse}
\end{figure}

While processing the \datasetname-Corpus, we discovered that several VR apps may choose the same document (referred to as \textit{reused policy document}) as their privacy policies even without modifying any content. Figure \ref{fig:finding_pp_reuse} shows the number of reused privacy policies and how often they are reused. It is observed for 3,521 apps providing privacy policies, 54.5\% (1919/3521) of them show the reuse behavior, with 687 unique policy documents being reused a total of 1919 times. 
Note that this statistic represents the minimum number of instances where privacy policies are reused, as some policies may also be used by other apps (including non-VR apps) beyond the scope of our \datasetname-Corpus. Two policies stood out, each with more than 15 reuses: Oculus's privacy policy document \cite{oculus-pp} (reused by 19 VR apps developed by Oculus) and Adobe's privacy policy document \cite{adobe-pp} (reused by 16 VR apps shared on Behance \cite{behance}, a social media platform owned by Adobe that focuses on showcasing creative work). We also identify the privacy policy of 17 different VR apps all re-direct to Microsoft's privacy policy \cite{microsoft-pp}.

\bheading{Privacy risks of reuse behavior of privacy policies.} 

A common pattern among these reused behaviors is that a certain developer publishes multiple VR apps and applies the same document as policy policies to all of them. In this scenario, the CUS claimed in the privacy policy typically is the union set of actual behaviors across all VR apps. As a result, despite the fact that two apps may have different data collection practices, users will see identical CUS in their reused privacy policy. 
For instance, \texttt{VZfit} (one of the top VR fitness apps with 611 ratings and rated 4/5) and \texttt{VZplay} (with 2.3k clicks and 16.2k views, rated 5/5 on Sidequest) are developed by VirZOOM Inc. and share the same privacy policy; 
however, they differ significantly in terms of their data collection practices, i.e., \texttt{VZfit} requires fine-grained location access and microphone permission from users, while \texttt{VZplay} does not request these permissions but accesses hand tracking data instead. Another example is Oculus, which not only functions as a platform but also operates as a developer. Within our \datasetname dataset, Oculus has released 19 official apps, all direct to the same privacy policy \cite{oculus-pp}. 
However, most of these apps display varying data collection behaviors. For example, \texttt{The World Beyond} (with 5.8k clicks and 19.5k views, rated 4.6/5 on Sidequest) collects microphone data and requests coarse-grained location access and hand-tracking data; whereas \texttt{Dear Angelica} (an interactive VR story developed by official Oculus Story Studio, ranked 9th among top-rated free VR apps with 281 ratings and an average rating of 4.35/5) does not have such practices.

From a commercial standpoint, it is cost-effective and legally sound for large companies or developers to maintain a single but comprehensive privacy policy that covers all possible data collection for their services and apps, regardless of the VR features or variations among different VR apps. However, this approach may lead to user confusion and breed mistrust.

\subsection{Vetting Results of Completeness}\label{sec:result:completeness}

\noindent \textbf{\textsc{Finding} \large 3}: \textbf{41.7\% VR privacy policies did not adequately inform users about their privacy rights and 65.9\% of relevant VR apps failed to address children's privacy concerns.}

As shown in Table \ref{tab:finding_parsing_ratio}, for all vetted privacy policies, only 34.7\% out of them provide all necessary 7 components described in Table~\ref{tab:vrchat_pp}. Over 81\% of privacy policies contain statements about data CUS, but less than 66\% of them provide other important components such as informing users' rights (58.3\%) to access, edit, and delete their personal data on the app's server, statements about data retention (57.1\%), and specific audiences (61.2\%). Some real-world examples of VR apps fail to provide complete privacy policies can be found on Table~\ref{tab:C2-violate} of Appendix~\ref{sec:example:violate}.
In the below, we present two case studies about policies without any valid component and improper handling of children's policy.

\bheading{Case study: privacy policy with no valid component}. Out of \datasetname-Corpus, only 34.7\% (1223/3521) of privacy policies provide ALL these components while surprisingly, 11.5\% (406/3521) provide NONE of these components. Upon manual inspection of 50 randomly selected privacy policies that lacked all these components mentioned above, we summarize three types of situations: (1) fake-redirection, where clicking on the privacy policy link redirects to the same page instead of the privacy policy content page, e.g., from \url{https://sloppystudio.com/} to \url{https://sloppystudio.com/#} on \texttt{Car Parking Simulator VR} (rated as \textit{Mostly Positive} in SteamVR); (2) ongoing privacy policy, where the page displays \textit{``Under Construction''} with no additional information, e.g., \texttt{Chandrayaan VR} 
; and (3) dummy privacy policy, where only one sentence is displayed, such as the privacy policy \cite{3d-event} 
\textit{``No user data is collected from the app manufacturer''} for \texttt{Las Vegas Helicopter Flight} and \texttt{Venice - Grand Canal}, both priced at 1.99 USD on Quest Store. While these privacy policies may maintain valid links, they do not provide any useful information.

\bheading{Case study: children's policy}. Given the popularity of immersive gaming experiences and the widespread use of VR devices by young children \cite{quest2_child_xmax, child-in-vr}, it is crucially important for VR app developers to address privacy concerns related to children in order to comply with children's data protection laws (e.g., COPPA \cite{ftc-coppa} 
). We focused on apps related to education, kids (children), or family genres and checked for specific statements regarding children's privacy. Out of 718 children-related apps we find, 65.9\% (473/718) do not mention anything about children's privacy in their policies. For instance, \texttt{Bogo} (a pet feeding game with 1326 ratings, ranked 8th among the top free VR apps with the highest number of ratings, and has an average rating of 4.2 out of 5), 
\texttt{Henry} (a storytelling app with 394 ratings and rated 4.0/5), and \texttt{Paper Birds} (an interactive story with 268 ratings and rated 4.4/5) are VR apps published on Quest and designed for users aged 3+. However, none of them contain specific statements about children's privacy in their policies. Specifically, the policy of \texttt{Paper Birds} only states that they \textit{``don’t collect any of your personal info at any time, ... have never received any legal or government demands for user information''}. A similar situation arises with two other representative VR apps, namely \texttt{Pets VR} (with 78.6k clicks and 264.4k views, rated 4.2/5 on Sidequest) and \texttt{Ultimate Fishing Simulator VR} (which has received 501 \textit{Very Positive} reviews on SteamVR). 

\begin{table}[t]
    \centering
    \footnotesize
    \caption{Ratio of privacy policies  with necessary components.}
    \begin{tabular}{ll|ll}
    \toprule
    Component       & Ratio(\%) & Component     & Ratio(\%) \\
    \midrule
    User Choice     & 65.6      & Data Security & 65.6      \\
    Data CUS & 81.1      & Policy Change & 64.2      \\
    User Rights     & 58.3      & Spec. Audience & 61.2      \\
    Data Retention  & 57.1      &    \textbf{All 7 Components}   &  \textbf{34.7}  \\ 
    \bottomrule
    \end{tabular}  \label{tab:finding_parsing_ratio}
\end{table}

\begin{figure}[t]
\centering
\subfigure[$PPG$ over privacy policies. \label{fig:finding_granularity_cdf}]{
\includegraphics[width=0.41\linewidth]{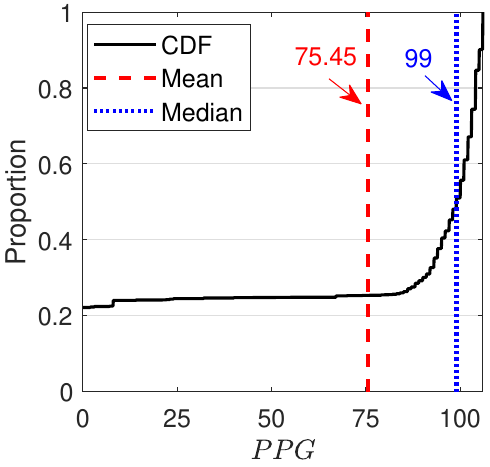}
}
\subfigure[Dataobj/entity granularity across CUS tuples. \label{fig:finding_granularity}]{
\includegraphics[width=0.54\linewidth]{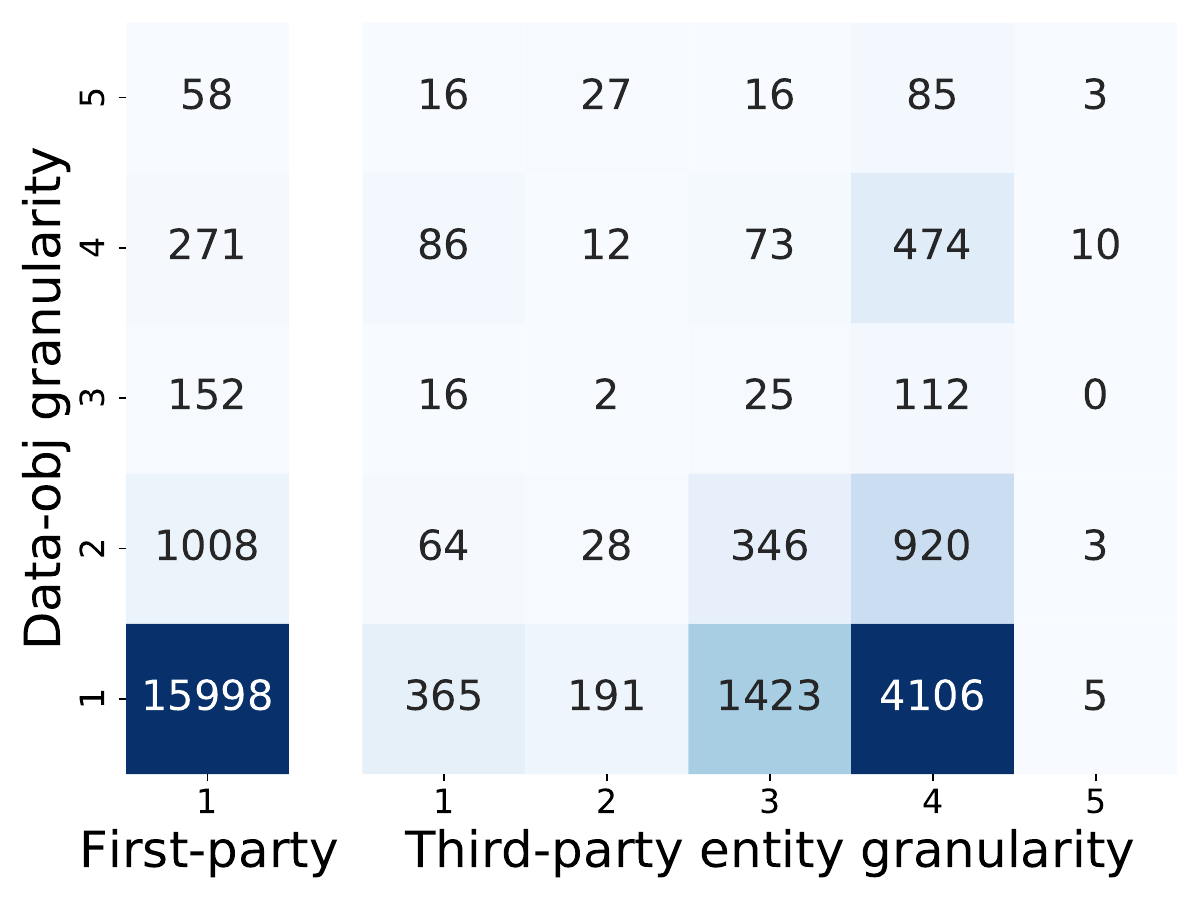}
}
\caption{Distributions of CUS tuple granularity and privacy policy granularity.\label{fig:granularity}}
\end{figure}

\subsection{Vetting Results of Granularity}
\label{subsec-deep-finding}

\noindent \textbf{\textsc{Finding} \large 4}: 
\textbf{Some VR privacy policies lack fine granularity in both data object disclosure and third-party specification.}

For totally 25,895 CUS tuples extracted from privacy policies in \datasetname-Corpus, we first calculate the $\mathit{PPG}$ of each privacy policy. Then we categorize all these CUS tuples into two groups: 17,487 first-party CUS tuples (where the entity is \textit{we} and the entity granularity equals 1) and 8,408 third-party CUS tuples (where the entity is explicitly or implicitly claimed as \textit{third-party}). We then calculate $\mathit{CTG}$ for each individual CUS tuple. The results are shown in Figure~\ref{fig:granularity}. 

\bheading{Granularity of privacy policy is coarse.} 
Figure \ref{fig:finding_granularity_cdf} displays the CDF of $\mathit{PPG}$ for all valid privacy policies. According to this figure, the inclusion of unspecified data objects in a privacy policy significantly increases the upper bound of claimed data objects. On average, a privacy policy will includes 75.4 data objects within the gap between its lower and upper bound. The median value is 99.0. Therefore, we can conclude that VR apps' privacy policies tend to use coarse-grained CUS statements rather than fine-grained ones.

\bheading{First-party CUS provides coarse-grained disclosure of data objects.}
It is observed from Figure~\ref{fig:finding_granularity} that 8.5\% (1,489/17,487) of first-party CUS tuples have a data object granularity higher than expectations (i.e., equal or larger than 2 in this study). Considering the definition of lower-bound, this indicates that these CUS are claimed \textit{without} any additional clarification or introduction in the privacy policy. Here are several typical examples:  18 privacy policies (including Intel's privacy policy \cite{intel-pp} for its VR app \texttt{Queerskins: Ark} in Viveport) fail to clarify the meaning of \textit{Biometric data}. Additionally, 79 privacy policies (including the well-known meditation and sleep VR app \texttt{Calm}) do not provide a clear definition for \textit{Health information}. Furthermore, 43 privacy policies, which include the popular game \texttt{Blobby Tennis} (with 48k clicks and 94.5k views) and the VR social platform \texttt{Cheerio} (with 2.6k clicks and 15.7k views), do not specify what is meant by \textit{Body Measurement}.

\bheading{Granularity of third-party CUS is even worse.}
27.6\% (2,318/8,408) of third-party CUS tuples do not specify the exact type of data object they collect, and surprisingly 93.5\% (7,861/8,408) of CUS tuples do not mention the exact company name of the third-party. The most common case is to refer to them as \textit{third-party} (4106). In other cases, they may be categorized as specific types of third parties such as \textit{ad network} (708) and \textit{platform provider} (578). Among the explicitly mentioned third parties, Google (186), Google Analytics (77), Unity (102), and Facebook (92) are the most prominent. It is worth noting that only 4.3\% (365/8,408) of these third-party CUS clearly identify both the company name of a third party and the specific types of data collected by that particular third party.

\begin{figure}[t]
    \centering
    \includegraphics[width=0.9\linewidth]{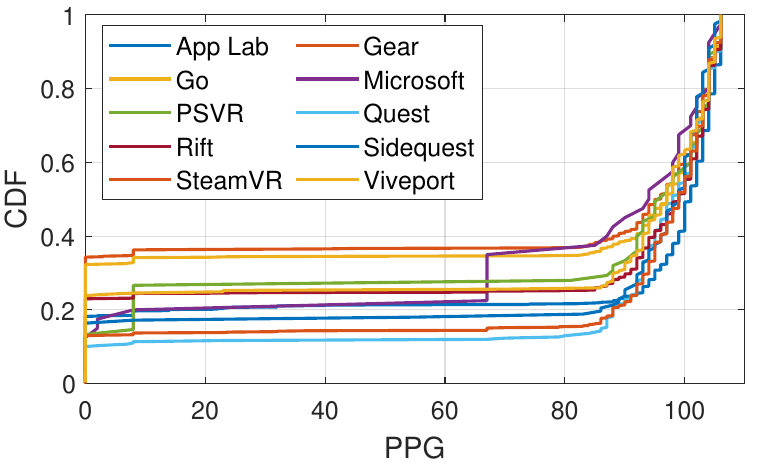}
    \caption{PPG over privacy policies on different VR platforms.}
    \label{fig:PPG-diff}
\end{figure}

\bheading{Granularity vetting results varies among different VR platforms.} Figure \ref{fig:PPG-diff} and Table \ref{tab:PPG-diff} display the differences of PPG results over privacy policies published on different VR platforms. Although there is little difference among their PPG-median values (from 94 to 101), in terms of PPG-mean value, apps in platform Gear averagely provide the most fine-grained privacy policies (with PPG-mean value as 63.57) while apps in platform Quest provide the least fine-grained granularity (with PPG-mean value as 86.89).

\begin{table}[t]
\caption{PPG mean and median of privacy policies on different VR platforms.}
\centering
\footnotesize
\label{tab:PPG-diff}
\begin{tabular}{ccc|ccc}
\toprule
Platform & PPG-mean & PPG-median & Platform & PPG-mean & PPG-median \\ \midrule
Gear & 63.57 & 96 & Rift & 75.37 & 99 \\
Go & 65.81 & 97 & App Lab & 80.23 & 101 \\
PSVR & 73.79 & 96 & Sidequest & 81.57 & 99 \\
Viveport & 74.11 & 97 & SteamVR & 84.88 & 99\\
Microsoft & 74.53 & 94 & Quest & 86.89 & 98\\ \bottomrule
\end{tabular}
\end{table}

\subsection{Vetting Results of Minimization}
\noindent \textbf{\textsc{Finding} \large 5}: \textbf{91.6\% of data objects and 85.1\% of VR apps exhibit at least one overbroad situation.}

\begin{figure*}[t]
    \centering
    \includegraphics[width=\linewidth]{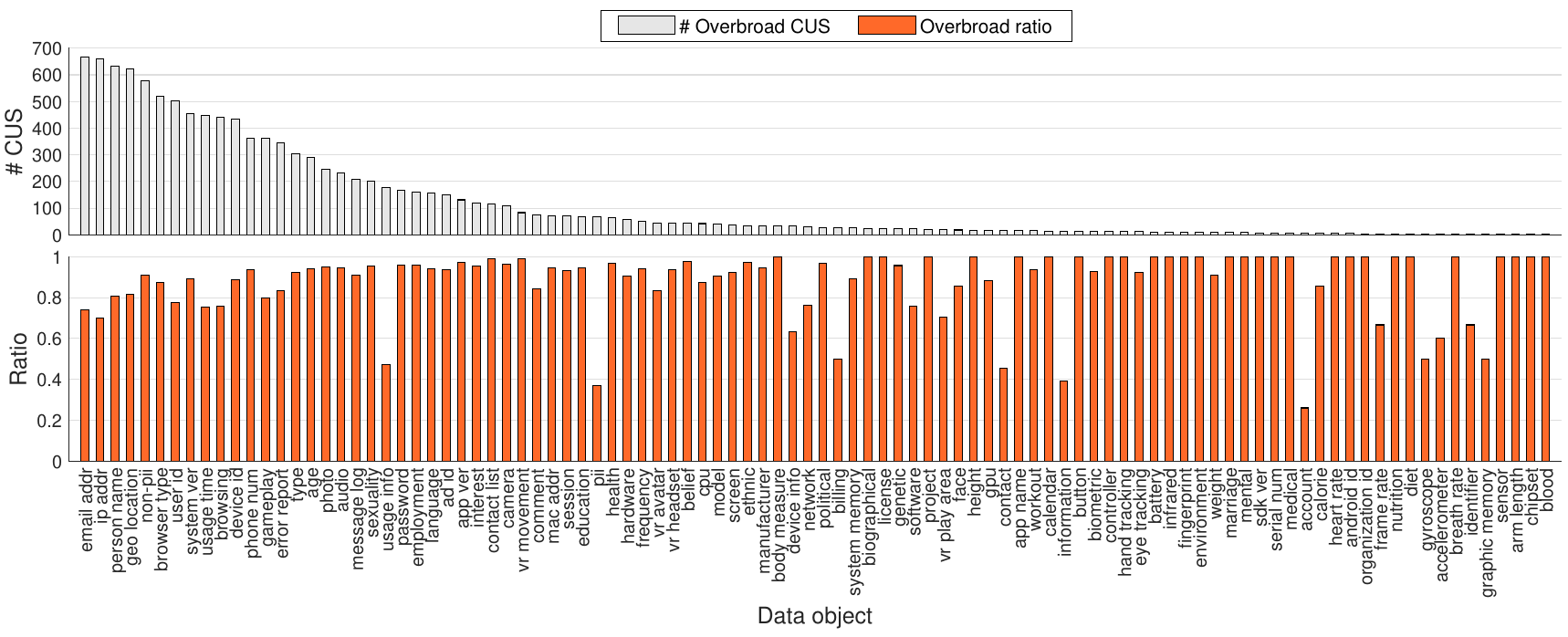}
    \caption{Distribution of overbroad data objects and their overbroad ratio.
    }
    \label{fig:finding_overbroad_lowerbound}
\end{figure*}

In total, we find counterparts for 2,033 VR apps,
and the minimization analysis results are shown in Figure \ref{fig:finding_overbroad_lowerbound}, which displays the identified overbroad data objects (i.e., surpassing their counterparts in CUS) along with their frequency and ratio. Out of all data objects in our ontology, we find that 91.6\% (98/107) are involved in at least one case of overbroad CUS. Similarly, out of all comparable VR apps (i.e., apps that we successfully find their counterparts), we find 85.1\% (1699/1997) had at least one overbroad data object in their CUS tuples. These findings demonstrate the prevalence of overbroad situations and suggest a need for stricter adherence to privacy policies' minimization principle.

As depicted in Figure \ref{fig:finding_overbroad_lowerbound}, 94.9\% (93/98) data objects have an overbroad ratio exceeding 0.5 (meaning a fifty-fifty chance of being overbroad when claimed in a VR app's privacy policy). The overbroad ratio of 65.3\% (64/98) of the data objects even exceeds 90\%, and there are 29 data objects that are consistently identified to be overbroad. It is worth noting that some VR-related data objects such as gameplay (362/79.7\%), audio information (231/94.7\%), camera information (109/96.5\%), and VR movement (83/98.8\%) have both a large number of overbroad CUS cases and high overbroad ratios.

\bheading{Case study of Finding 5: All-encompassing privacy policy of large company used on VR app.}
During our evaluation of the overbroad situation in privacy policies, we discovered that some VR apps link to a parent company's or association's privacy policy (referred to as an \textit{all-encompassing} privacy policy). Note that this phenomenon is different from what we refer to as \textit{reused} privacy policies in \textbf{Finding 2}, which are shared among different VR apps and include the union set of these VR apps' CUS. The \textit{all-encompassing} privacy policies, instead, include all CUS requirements for the parent company's business affairs, without making any modifications or adding specific statements that apply to the traits of the VR app. Consequently, these privacy policies end up claiming more data objects than what the VR app actually collects. As a result, our evaluation highlights these policies as failing to meet the minimization criterion.

Taken \texttt{Qantas VR} (published in SteamVR) and \texttt{The Emirates VR Experience} (published in Rift) as examples, they claim to collect highly sensitive information such as ethnicity, beliefs, and passport details - unnecessary and impractical for flight experience VR apps. It turns out that these two apps link their privacy policy to the policy of their respective parent companies (Qantas and Emirates), where large amounts of personal information are claimed to be collected. Similar situations occur with \texttt{Lusail Stadium VR Experience} (published in Go and Gear) linking to Qatar 2022 FIFA World Cup's privacy policy, claiming to collect user passport information; and \texttt{BYU Virtual Campus} (published in SteamVR) linking to Brigham Young University's (the largest church university in the USA) privacy policy, claiming to collect user education, employment, belief, and mental health information. Other real-world examples can be found in Table~\ref{tab:C4-violate} of Appendix~\ref{sec:example:violate}.

\bheading{Minimization vetting results varies among different VR platforms.} Table \ref{tab:Mini-diff} displays the differences of minimization results of those privacy policies published on different VR platforms. Since the sizes and the sources for searching for counterparts are different for different VR platforms, the number of VR apps that are \textit{comparable} for minimization vetting would be different. Among all the platforms, PSVR has the lowest overbroad ratio, indicating that VR apps on PSVR have the \textit{most minimized} privacy policies. In contrast, MicrosoftMR has the highest overbroad ratio, with 28 out of 29 comparable VR apps deemed to be overbroad.

\begin{table}[t]
\caption{Minimization vetting results of privacy policies on different VR platforms.}
\centering
\footnotesize
\label{tab:Mini-diff}
\begin{tabular}{cccc}
\toprule
Platform & \#Comparable & \#Overbroad & \begin{tabular}[c]{@{}c@{}}$\frac{\#Overbroad}{\#Comparable}$\end{tabular} \\
\midrule
PSVR & 62 & 46 & 0.742 \\
App Lab & 243 & 183 & 0.753 \\
Sidequest & 20 & 16 & 0.800 \\
Viveport & 196 & 168 & 0.857 \\
SteamVR & 685 & 589 & 0.860 \\
Rift & 807 & 696 & 0.862 \\
Go & 516 & 456 & 0.884 \\
Gear & 481 & 426 & 0.886 \\
Quest & 252 & 230 & 0.913 \\
Microsoft & 29 & 28 & 0.966 \\
\bottomrule
\end{tabular}
\end{table}

\subsection{Vetting Results of Consistency}

\noindent \textbf{\textsc{Finding} \large 6}: \textbf{Inconsistencies between actual code behavior and privacy policies are common in VR apps, particularly with regard to VR-related data objects.}

This analysis is based on \datasetname-APK. We first observe that the issue of poor accessibility of privacy policies, as mentioned in \textbf{Finding 1} about Sidequest, still persists. Out of all 1,096 apps, 810 (662 of which come from Sidequest) do not provide the required privacy policy. Furthermore, the APK files of their apps exhibit evidence of sensitive permissions or behaviors, as indicated in the first column of Table \ref{table:evaluation-consistency}. This could potentially breach the relevant privacy law (if applicable) because there is a discrepancy between the privacy policy (NULL here) and the actual behavior of the app.

\begin{table}[t]
\centering
\footnotesize
\caption{Evaluation results of consistency analysis. The cumulative results are shown in the Total row. For the \#App, they are further de-duplicated based on names.}
\label{table:evaluation-consistency}
\resizebox{\linewidth}{!}
{\begin{tabular}{ccccc}
\toprule\
\multirow{2}{*}{Data object} & w/o pp & \multicolumn{3}{c}{w/ pp} \\ \cline{3-5}
 & \# App & \# App & \# / \% Vague& \# / \% Inconst.\\ \midrule
network & 782& 286& 221 / 77.3\%& 39 / 13.6\%\\
geo location & 732& 262& 149 / 56.9\%& 35/ 13.4\%\\
device info & 703& 252& 165 / 65.5\%& 36 / 14.3\%\\
audio & 412& 118& 67 / 56.8\%& 11 / 9.3\%\\
camera & 164& 51& 42 / 82.4\%& 7 / 13.7\%\\
\rowcolor[HTML]{EFEFEF} hand tracking & 128& 37& 31 / 83.8\%& 5 / 13.5\%\\
billing & 71& 50& 35 / 70.0\%& 5 / 10.0\%\\
account & 58& 27& 19 / 70.4\%& 5 / 18.5\%\\
usage info & 45& 20& 6 / 30.0\%& 1 / 5.0\%\\
\rowcolor[HTML]{EFEFEF} vibrator & 26& 12& 12 / 100.0\%& 
 --\\
ad id & 26& 12& 10 / 83.3\% & 1 / 8.3\%\\
\rowcolor[HTML]{EFEFEF} vr play area & 5& 7& 7 / 100.0\%& --\\
\rowcolor[HTML]{EFEFEF} infrared & 2& 7& 7 / 100.0\%& --\\
\rowcolor[HTML]{EFEFEF} eye tracking & 2& 6& 5 / 83.3\%& --\\
 contact& 2& -& -&-\\
\rowcolor[HTML]{EFEFEF} pupil distance & 1& 6& 6 / 100.0\%& --\\
\rowcolor[HTML]{EFEFEF} face & 1& 5& 4 / 80.0\%& --\\
accelerometer & 1& 2& 2 / 100.0\%& --\\
\rowcolor[HTML]{EFEFEF} body measure & 1& 2& 2 / 100.0\% & --\\
gyroscope & 1& 2& 2 / 100.0\% & --\\
error report & 1& 2 & 2 / 100.0\% & --\\
\rowcolor[HTML]{EFEFEF} vr movement & 1& 2& 2 / 100.0\% & --\\
\rowcolor[HTML]{EFEFEF} environment & 1& 4& 4 / 100.0\% & --\\
 battery& 1& 1& 1 / 100.0\% & --\\
 \rowcolor[HTML]{EFEFEF} heart rate& 1& 1& 1 / 100.0\% & --\\
 \rowcolor[HTML]{EFEFEF} fingerprint& 1& 1& 1 / 100.0\% & --\\
 \rowcolor[HTML]{EFEFEF} biometric& 1& 1& 1 / 100.0\% & --\\ \midrule
\textbf{Total} & \textbf{810}& \textbf{286}& \textbf{809 / 68.4\%}& \textbf{147 / 12.4\%}\\ \bottomrule
\end{tabular}}
\end{table}

Then we analyze the inconsistency of those apps (a total of 286) with both privacy policy and apk file available and the results are shown in the right part of Table \ref{table:evaluation-consistency}. We found that 85.3\% (244/286) of these apps vaguely claim their code behavior, while 15.7\% (45/286) do not disclose their code behavior in the privacy policy. From the perspective of per code behavior, the overall inconsistency ratio is 12.4\% (147/1183), with the majority of them (74.8\%, or 110/147) related to three common data objects: network information, geo-location, and device information. For camera and hand tracking data objects, although the total number of inconsistencies is not significant, the relative inconsistency ratio in these data objects is higher than average, with 13.7\% and 13.5\% respectively.
It is worth noting that we once again confirm the coarse granularity of CUS in VR app privacy policies. We have observed that 68.4\% CUS practices in VR app code are not explicitly mentioned in their corresponding privacy policy. Instead, they are covered by a vague CUS statement with coarse granularity. Surprisingly, over half of the data objects (14/27) listed in Table \ref{table:evaluation-consistency} show vagueness across all their occurrences in VR apps. Among these, 71.4\% (10/14) are due to highly VR-related data objects such as \textit{vr play area} and \textit{pupil distance}.

\bheading{Current platform pay insufficient attention to policy's vagueness and inconsistency.} Towards this phenomenon, we might find some relevant explanations on the official Meta website. Meta considers a privacy policy necessary in their VRC (Virtual Reality Check) and conducts a detailed Data Use Checkup (DUC) when reviewing an app. However, we discover that it informs developers \textit{``It does not need to provide the explicit data (e.g. Username, Profile picture, leaderboards) and can reference it in abstract''} in their second and third criteria for the privacy policy \cite{meta-guideline, meta-guideline2}. This suggests that Meta may encourage developers to use coarse-grained data collection statements even if both Meta and the developers are aware of the specific data types collected from users.

\section{Discussions}
\label{sec-limitation}

\bheading{Impact of privacy policy correctness on vetting results.} Except for the consistency and availability criteria, the other three criteria (completeness, granularity and minimization) are designed to vet the \textit{quality} rather than the \textit{correctness} of VR app's privacy policy. For instance, consider a maliciously curated privacy policy. Assume this privacy policy: 1) includes a \textit{Data Security} component but its VR app does not implement the claimed data protection measures, and 2) specifically states that it collects very limited types of user data, while in reality, the VR app attempts to collect as much user data as possible. As a result, this incorrect privacy policy would pass all the completeness, granularity and minimization vetting criteria and be considered as a \textit{high-quality} privacy policy, even though the VR app's actual behavior does not align with it.
Although our consistency criteria partially address this issue, the challenge of verifying the correctness of all components of a privacy policy remains an open problem.

\bheading{Limited collection of privacy policy and package file.}
For privacy policies, we only considered easily accessible ones in Section \ref{subsec-system-data-collection}, thus may ignore the cases where the privacy policy was not readily displayed on the homepage but could be accessed by, for example, adding a suffix like \textit{\url{/privacy-policy}} to the URL. We may also miss those privacy policies that use keywords (e.g., ``statement'', ``notice'', ``legal'', ``terms'', ``agreement'', ``disclaimer'', and ``policy'') other than ``privacy'' on their homepage. Out of randomly selected home pages of 1k VR apps, we identified 14 privacy policies belong to this situation.
For package files, we encountered significant network failures when attempting to download them. A simple retry can increase the number of successfully downloaded apps from 350 to 691. Therefore, we repeated our method three times and obtained the results in Section \ref{subsec-dataset-des}. We leave enhancing the data collection method's robustness for future work.

\bheading{Context-aware CUS tuple extraction.} Our CUS tuple model only considers the entity and data object of a CUS, ignoring other contexts. For instance, it is common to put a CUS sentence like \textit{``when you $\langle$condition$\rangle$, $\langle$entity$\rangle$ may collect your $\langle$data obj$\rangle$ to $\langle$purpose$\rangle$''} where extended CUS tuple can provide context integrity (CI) \cite{shvartzshnaider2019going, apthorpe2019evaluating} for privacy handling. This extended CUS tuple can help analyze privacy policy in a more holistic approach. However, automatically extracting context with high accuracy remains an open problem, and we will leave it as future work.

\bheading{Incomplete coverage of VR apps' behaviors.} Firstly, the analysis is limited to apps that satisfy both StandaloneVR and Free conditions.
Secondly, there are limitations in the coverage of app behaviors as sensitive permissions and APIs may include deprecated or unreachable dead code, resulting in false positives. Additionally, data objects covered by permissions or APIs are incomplete compared to the full data object ontology. Certain sensitive data categories such as mental health and preferences cannot be identified merely through code, but rather require a combination of other data types for inference. 
It may need to explore the intricate relationships between data types in VR scenarios and develop more precise methods for modeling VR app behaviors. We believe that an automatic tool capable of analyzing VR apps both statically and dynamically is necessary to cover as many VR data collection behaviors as possible. A promising direction would be developing a VR device simulator that can run VR APPs and inject inputs that simulate human interactions.
\section{Related works}
\label{sec-related}

\bheading{Privacy issues in VR.} The privacy concerns of VR apps stem from their immersive experience and the ability to collect sensitive data \cite{jia2017ethical, de2019security, happa2021privacy}. Various sensor data, such as motion, eye tracking, hand tracking, indoor tracking, facial capture, and body measurement \cite{bye2019ethical,buck2021privacy}, can be used to infer users' health or profile information with high accuracy \cite{bailenson2018protecting,nair2022exploring}. A semi-structured interview with 20 VR users and developers \cite{adams2018ethics} explore the privacy perception in the VR community and co-design a set of ethics regarding VR content, which could potentially serve as industry-wide standards.

\bheading{CUS extraction of privacy policies.} There are two main technical approaches to extracting CUS statements from privacy policies. First is the rule-based extraction (RBE) solution proposed in PolicyLint \cite{policylint} (also adopted by PoliCheck \cite{policheck} and OVRSeen \cite{ovrseen}) and PoliGraph \cite{poligraph}. PolicyLint utilizes data-entity-dependency (DED) patterns to identify and extract CUS tuples from sentences, but it has a relatively low recall rate (<30\%) due to limited pre-defined patterns. PoliGraph also extracts CUS tuples based on several rules. To improve the recall rate, it has to decompose this task into 5 relation-annotator tasks, including collection, subsumption, and coreference relations, and combines the results of these tasks on a graph structure. 
Second is the end-to-end extraction, as proposed by PI-Extract \cite{bui2021automated}. However, the CUS tuple extracted through this method lacks accurate entity information and only separates them into first-party or third-party collections.

\bheading{Privacy policies in other domain.} In addition to websites and mobile phone apps, privacy policies are required in many other domains. Perez \textit{et al.} \cite{perez2018review} and Yu \textit{et al.} \cite{yu2021whether} have examined the privacy policies of IoT devices and their consistency with actual app behaviors. Manandhar \textit{et al.} \cite{manandhar2022smart} did a thorough survey of privacy policies in the Smart Home domain, identifying 17 findings that impact millions of users. Bui \textit{et al.} \cite{bui2023detection} proposed ExtPrivA, an automatic tool that detects inconsistencies between browser extensions' data collection and their privacy policies. Other studies focus on specific categories of applications such as menstrual apps \cite {shipp2020private}, or money services \cite{bowers2017regulators}. Trimananda \textit{et al.} developed OVRSeen \cite{ovrseen} to audit network traffic and privacy policies in Oculus VR. However, in addition to the limitation of CUS extraction discussed above, OVRSeen  covers only 150 VR apps and ignores other criteria for vetting a policy except for consistency. 

\bheading{Privacy policy generator.} Drafting a privacy policy can be complex, which is why there are tools available to generate policies for Android apps \cite{yu2015autoppg, yu2016toward}, iOS apps \cite{zimmeck2021privacyflash}, and smart home apps \cite{li2021toward}. Google announced Checks \cite{checks} to help developers draft and rectify their privacy policies according to relevant laws. However, these tools face challenges in extracting reasons for data collection practices from source code. Besides, some components like user rights cannot be inferred from app source code alone. 

\section{Conclusion}
\label{sec-conclusion}

This study proposes \sysname, the first large-scale comprehensive privacy policy vetting system for VR apps. \sysname collects 11,923 different VR apps' meta-info and analyzes 3,521 valid privacy policies 
 (from 10 mainstream VR platforms) 
as well as 1,096 VR apps' package files 
 (from 3 VR platforms) 
based on the following 5 criteria: availability, completeness, granularity, minimization, and consistency. Our findings expose the significant mishandling and disregard for privacy within the current VR ecosystem. To increase awareness of the VR community and facilitate future analysis of VR privacy policies, we open-source \sysname system as well as our research findings on \url{https://github.com/kalamoo/PPAudit}.

\section*{Acknowledgements}
The authors from Shanghai Jiao Tong University were partially supported by the National Natural Science Foundation of China (No. 62325207, 62132013, 62302298, 62332013), Young Elite Scientists Sponsorship Program by CAST (YESS20230589), and Startup Fund for Young Faculty at SJTU (23X010502192). The authors from Xidian University were partially supported by the National Natural Science Foundation of China (No. 62302362) and the Key Research and Development Programs of Shaanxi under Grants S2024-YF-YBGY-1540. 
Yichang Xiong and Xiaokuan Zhang were partially supported by a seed grant from the CAHMP center at George Mason University.

\bibliographystyle{ACM-Reference-Format}
\bibliography{reference}

\appendix

\begin{algorithm}[t]
    \caption{LowerBound}
    \footnotesize
    \label{alg:lower-bounds}
    \KwIn{All data objects $D$ in extracted CUS tuples from a privacy policy and data ontology $O$ (whose node set is $V$)}
    \KwOut{Lower bounds $LB$ of $D$.}

    \tcp{Calculate NodeGranularity (NG) of each node and store them accordingly.}
    $G \gets []$ \; 
    $maxNG \gets 0$ \;
    \For{node $n$ in $V$}{
        ng = \text{NG}(O, n) \;
        update maxNG \;
        append $n$ to $G_{\text{ng}}$ \;
    }

    \tcp{Initialize Lower bounds as all leaves in $D$}
    LB $\gets G_0$ \;

    \tcp{Iterate each node by their NG score}
    \For{i = 1 to maxNG}{ 
        NodesOfInterest = $D \cap G_i$ \;
        NodesToDelete = \{\ \} \;
        \For{node $u$ in NodesOfInterest}{
            \For{node $v$ in LB}{
                \If{HasPath($G$, $u$, $v$)}{
                    add node $u$ to NodesToDelete \;
                    \textbf{break} \;
                } 
            }
        }
        LB = (LB $\cup$ NodesOfInterest) $\setminus$ NodesToDelete \;
    }
\end{algorithm}

\begin{algorithm}[t]
    \footnotesize
    \caption{UpperBound}
    \label{alg:upper-bounds}
    \KwIn{All data objects $D$ in extracted CUS tuples from a privacy policy, data ontology $O$}
    \KwOut{Upper bounds $UB$ of $D$.}
    UpperBounds $\gets$ $D$ \;
    \For{node $n$ in $D$}{
        UB = UB $\cup$ Descendants($O$, $n$)\;
    }
\end{algorithm}

\begin{table*}[t]
    \centering
    \footnotesize
    \caption{Legal articles related to vetting 5 criteria across various regions.}
    \begin{tabular}{ccllll}
    \toprule
    \multicolumn{3}{c}{Requirements}                                & GDPR                                           & CCPA                                        & PIPL       \\ 
    \midrule
    \multirow{8}{*}{PP}  & \multicolumn{1}{l}{C1} & Availability    & 5.1(a,b), 12.1                                 & 1798.100                                    & 1          \\
                         & \multirow{7}{*}{C2}    & User Choice     & 5.1(a), 6, 7, 8, 9, 12-22, 49                  & 1798.120, 999.306, 999.308(c)(3)            & 15, 24, 44 \\
                         &                        & Data Collection & 5.1(a-c), 4-11, 19, 24-25, 28-30, 33-29, 44-49 & 1798.100, 1798.115, 999.305, 999.308        & 17, 23     \\
                         &                        & User Rights     & 5.1(a,d), 12-22                                & 1798.105, 1798.106, 1798.110, 999.308(c)(2) & 45-47      \\
                         &                        & Data Retention  & 5.1(e), 5, 25, 30                              & 1798.125                                    & 17         \\
                         &                        & Data Security   & 5.1(f), 6, 32-34                               & 999.313, 999.326                            & 57         \\
                         &                        & Policy Change   & 5.1(a)                                         & -                                           & 17         \\
                         &                        & Spec. Audience  & 5.1(a), 8                                      & 999.330-332                                 & 28,31      \\ 
    \midrule
    \multirow{3}{*}{CUS} & \multicolumn{1}{l}{C3} & Granularity     & 5.1(a-b), 24-29                                & 1798.100                                    & 7          \\
                         & \multicolumn{1}{l}{C4} & Minimization    & 5.1(c)                                         & 1798.140(e)                                 & 6          \\
                         & \multicolumn{1}{l}{C5} & Consistency     & 5.1(a-e)                                       & 1798.100                                    & 7          \\ 
    \bottomrule
    \end{tabular}
    \label{tab:privacy_law}
\end{table*}

\section{VR synonyms and ontologies.}
\label{apdx-cus-pipeline}

To alleviate the manual burden of checking a large number of un-terminologized phrases, we propose an approach based on the insight that synonyms have similar semantics and are therefore clustered in the embedding space.

\begin{figure*}[t]
    \centering
    \subfigure[VR data ontology. \label{fig:data-onto}]{
        \includegraphics[width=0.48\linewidth]{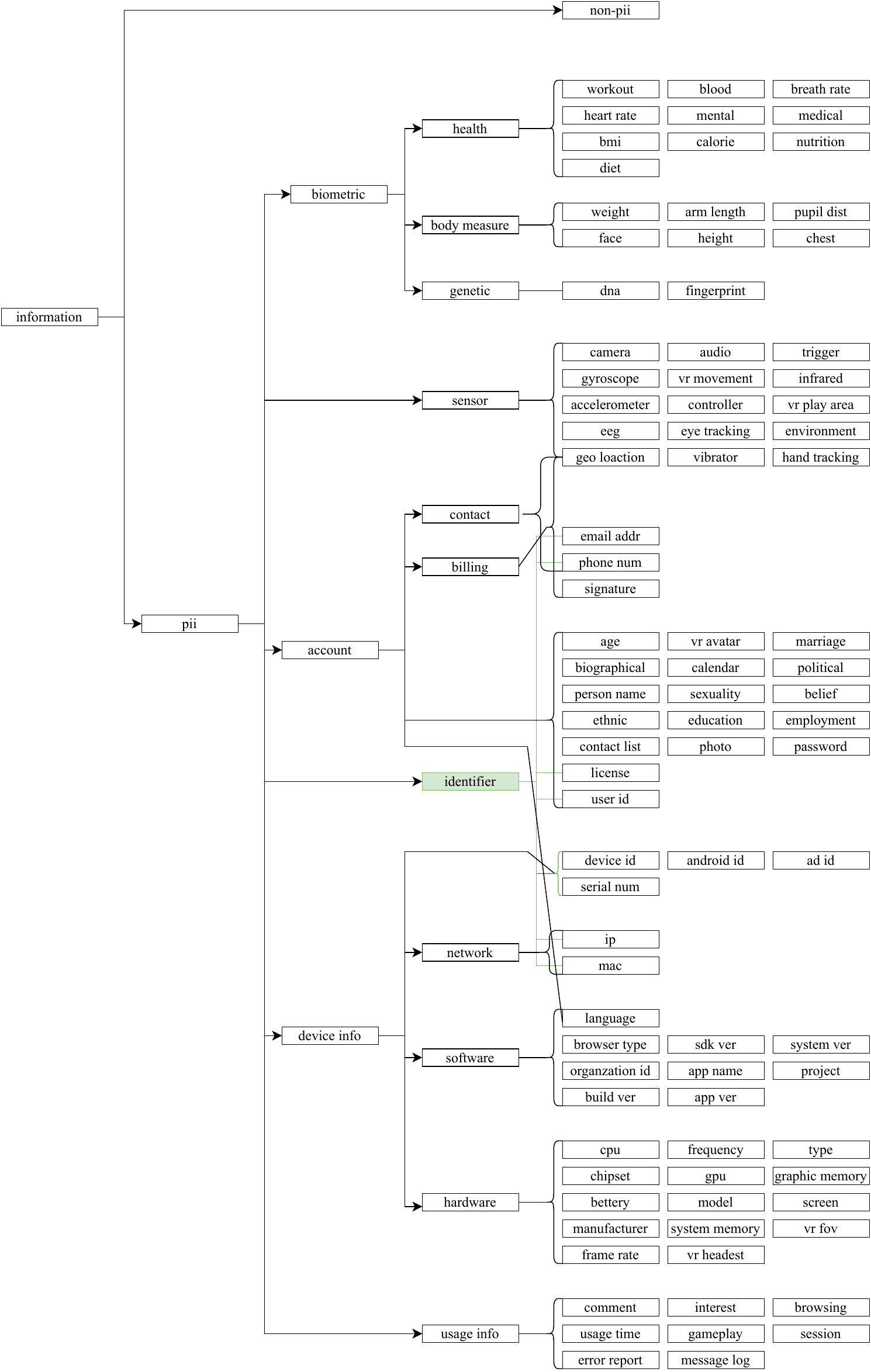}
    }
    \subfigure[VR entity ontology. \label{fig:entity-onto}]{
        \includegraphics[width=0.48\linewidth]{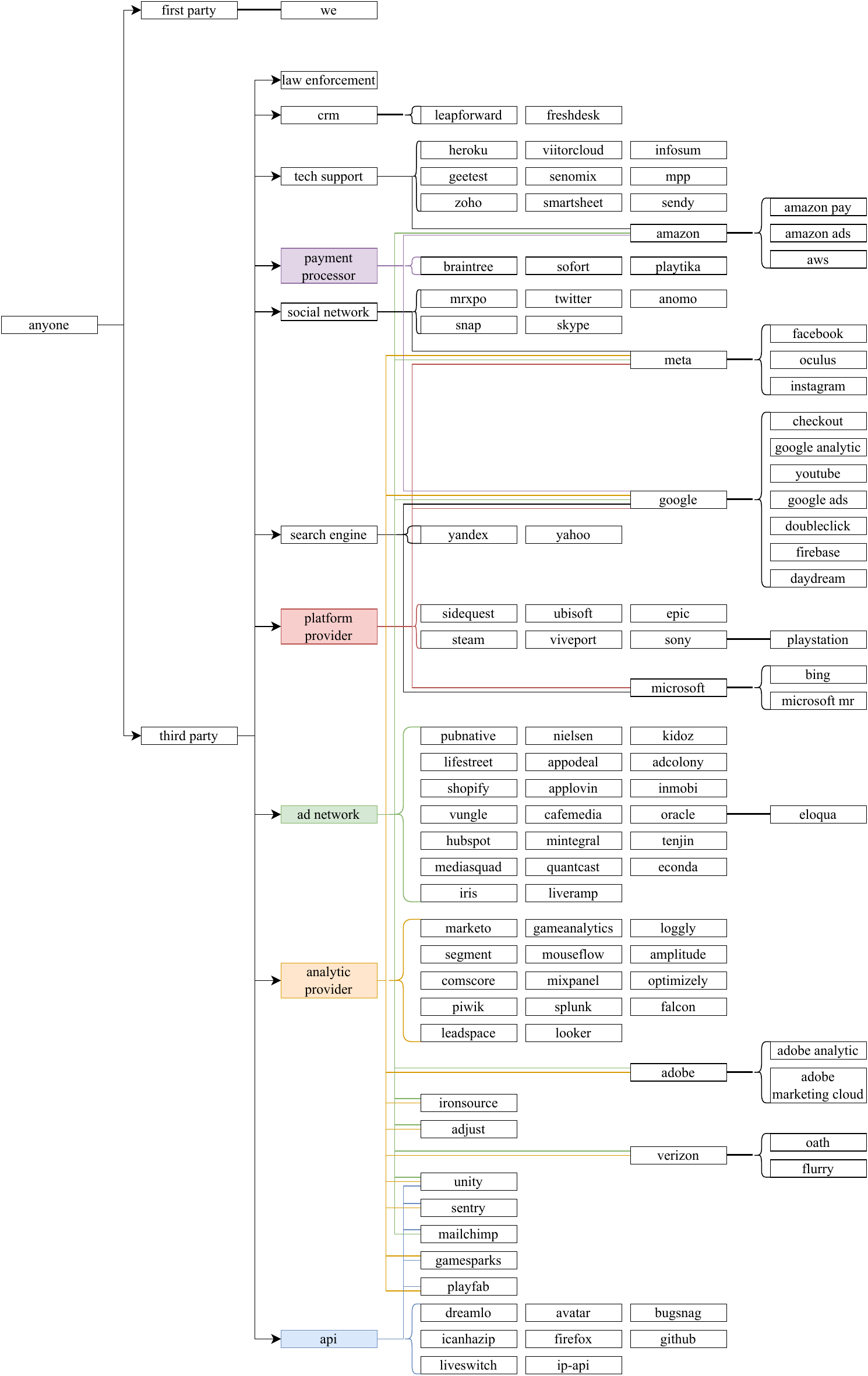}
    }
    \caption{Extended VR data ontology and entity ontology. To be concise, some arrows are omitted and the left bracket indicates edges from the parent node to all nodes inside.}
    \label{fig:onto}
\end{figure*}

To process data object phrases, we first utilize a BERT-based sentence embedding model to map all phrases to the semantic embedding space. During this phase, any phrases that are within a threshold distance (0.8 in our study, i.e., the median context similarity of the OVRSeen synonyms file) are added to synonym lists. For remaining unterminologized phrases, we iteratively spot new clusters in embedding space and determine whether they can be included in the VR data ontology and where they should be placed. During this process, we assume a broadly inclusive position for what personal information counts as PII only if they do not claim to be non-PII, as prior work \cite{henriksen2016re} has shown identity can be reconstructed from a variety of information types. As a result, we obtain an extended VR data ontology with 107 nodes and corresponding synonym lists containing 8,042 distinct data object phrases (See Figure \ref{fig:data-onto}). 

When dealing with entity phrases, we consider two situations: entity category and company name. Entity categories such as \textit{payment processor} describe the functionality of an entity and have semantic meaning. For instance, \textit{credit card company} and \textit{billing company} are close in embedding space and can be considered synonyms for \textit{payment processor}. On the other hand, company names like \textit{checkout} and \textit{braintree}, though both are payment processors, share no semantic similarity. Therefore, we identify synonyms of the company names through keyword matching since their names are context-free and remain identical. When handing implicit first-party entities (i.e., using the company name instead of \textit{we}), we follow the guidelines outlined in \cite{ovrseen}, and also take into account (1) the app name, (2) publisher or developer name, and (3) domain names found in both homepage links and privacy policy links. 
In total, we obtain an extended VR entity ontology containing 117 nodes along with synonym lists containing 1,663 distinct entity phrases for each category of entities (See Figure \ref{fig:entity-onto}).

\section{Legal basis of five criteria}
\label{apdx-law}

Some legal policies (e.g., GDPR, CCPA, PIPL) related to our vetting criteria are listed in Table \ref{tab:privacy_law}.

\section{Examples of VR Apps Violating Vetting Criteria}
\label{sec:example:violate}

Tables~\ref{tab:C1-violate},~\ref{tab:C2-violate}, and \ref{tab:C4-violate} show some typical VR apps which violate the criteria of availability, structural completeness, and minimization requirement.

\begin{table*}[t]
    \centering
    \caption{Partial vetting results of C1-Availability. Note that the full version of the vetting results and the privacy report (aggregate by every single privacy policy) will be uploaded to our project page. Popularity represents \# Rating (or Downloads) of this app. Rank represents the rank of \# Rating (or Downloads) among all apps in the platform where the app is published.}
    \footnotesize
    \begin{tabular}{lllllll}
    \toprule
        App Name & Platform & Developer & Price & Popularity & Rating & Rank\\
    \midrule
\texttt{Vrhealth Portal}        & Go, Gear  & xrhealth ltd        & Free    & 48         & --     & \textasciitilde 500 \\
\texttt{Balls and Hoops}        & Go, Gear  & delightly creative  & Free    & 26         & --     & \textasciitilde 600 \\
\texttt{Asteroids VR}           & Microsoft & baobab studios      & Free    & 12         & --     & 9                                  \\
\texttt{Theroseandi}	& Microsoft & penroses studios & Free & 9 & -- & 11 \\
\texttt{Verto Studio VR}        & Microsoft & --                  & \$29.99 & 1          & --     & 45                                 \\
\texttt{Atrain Express}         & PSVR      & komodo co ltd       & \$49.99 & -- & -- & -- \\
\texttt{Afterlife} & PSVR & signal space inc & \$1.99 & -- & -- & -- \\
\texttt{Archangel} & PSVR & skydance interactive llc & \$29.99 & -- & -- & -- \\
\texttt{Attack on Quest}        & Sidequest & --                  & Free    & 846.9k     & 4.55   & 3                                  \\
\texttt{Physics Playground}     & Sidequest & --                  & Paid    & 613.5k     & 4.48   & 5                                  \\
\texttt{Bmbf} & Sidequest & -- & Free & 590.7k & 4.03 & 6 \\
\texttt{Devour}                 & SteamVR   & straight back games & \$4.99  & 36,578     & --     & 10                                 \\
\texttt{Hand Simulator}         & SteamVR   & hfm games           & \$1.99  & 27,026     & --     & 17                                 \\
\texttt{Soulofsango}            & Viveport  & visense             & Free    & 152        & 4.2    & 3                                  \\
\texttt{Mercenary 3dof Edition} & Viveport  & kukrgame            & \$8.99  & 70         & 4.4    & 15 \\
        ... & ...  & ...            & ...  & ...   & ...  & ...\\
    \bottomrule
    \end{tabular}
    \label{tab:C1-violate}
\end{table*}

\begin{table*}[t]
    \centering
    \caption{Partial vetting result of C2-Completeness.}
    \footnotesize
    \begin{tabular}{llllllll}
    \toprule
        App Name & Platform & Developer & Price & Missing component(s) & Popularity & Rating & Rank\\
    \midrule
\texttt{Vspeedway}           & App Lab          & --                        & Free    & Data Retention, User Choice     & 390.2k     & 4.47   & 11                                 \\
\texttt{Tea for God}         & App Lab          & --                        & Paid    & Data Retention, Policy Change   & 220.3k     & 4.75   & 17                                 \\
\texttt{Samsung Befearless}  & Gear             & samsungbefearless         & Free    & Data Security, Spec. Audiences  & 4,376      & --     & 28                                 \\
\texttt{Robert Rodriguezs}   & Gear             & stxsurreal                & \$9.99  & User Choice, User Rights        & 246        & --     & \textasciitilde 200 \\
\texttt{Battle Talent}       & App Lab, SteamVR & saimon                    & Paid    & User Choice                     & 421.6k     & 4.71   & 8                                  \\
\texttt{Walk the World}      & Microsoft        & localjoost                & Free    & Data Security, User Choice      & 17         & --     & 6                                  \\
\texttt{Hello Mars}          & Microsoft        & voox singapore pte ltd    & \$4.99  & Policy Change, Data Collection  & 15         & --     & 7                                  \\
\texttt{A Fishermans Tale}   & PSVR             & vertigo games bv          & \$19.99 & Data Security, Data Retention   & --         & --     & --                                 \\
\texttt{Prey Digital}        & PSVR             & bethesda softworks        & \$39.99 & Spec. Audiences, User Rights    & --         & --     & --                                 \\
\texttt{Alcove} & Go, Quest & aarp innovation labs & Free & Data Retention & 160 & -- & \textasciitilde 300 \\
\texttt{Asgards Wrath} & Rift & sanzaru & \$39.99 & Data Collection, Spec Audiences & 3,550 & -- & 12 \\

\texttt{Affected the Manor}  & PSVR, Quest      & fallen planet studios ltd & \$9.99  & User Access                     & 846        & --     & \textasciitilde 150 \\
\texttt{Face Your Fears}  & Rift  & turtle rock studios  & Free  & Data Security, Data Collectio & 1,433 & -- & 28 \\
\texttt{Blade Sorcery Nomad} & Quest            & warpfrog                  & \$19.99 & Data Security                   & 19,964     & --     & 3                                  \\
\texttt{Triton VR}           & Sidequest        & --                        & Paid    & Spec. Audiences, Data Retention & 112.8k     & 4.14   & 46                                 \\
\texttt{Tabletop Simulator}  & SteamVR          & berserk games             & \$19.99 & Data Retention                  & 32,251     & --     & 13                                 \\
\texttt{The Rose and I}      & Viveport, Rift   & penrose studios           & Free    & Data Security, Data Retention   & 22         & 4.0    & 122 \\
        ... & ...  & ...            & ...  & ...   & ...  & ...\\
    \bottomrule
    \end{tabular}
    \label{tab:C2-violate}
\end{table*}

\begin{table*}[t]
    \centering
    \caption{Partial vetting result of C4-Minimization.}
    \footnotesize
    \begin{tabular}{llllllll}
    \toprule
        App Name & Platform & Developer & Price & Overbroad Data & Popularity & Rating & Rank \\
    \midrule
        \texttt{Paradiddle} & App Lab & -- & Paid & email, age, sexuality, geo-location & 183.7k & 4.57 & 26 \\
        \texttt{Gladius} & App Lab & -- & Paid & email, person name, usage info & 122.4k & 4.65 & 40 \\
        \texttt{Battle Talent} & App Lab, SteamVR & saimon & Paid & geo-location, person name & 421.6k & 4.71 & 8 \\
        \texttt{Star Wars Droid Repair Bay} & Gear & ilmxlab	& Free & usage info & 1,065 & -- & \textasciitilde 100 \\
        \texttt{Smash the Beats} & Go & nivision & \$7.99 & pii & 448 & -- & \textasciitilde 150 \\
        \texttt{Rilix Coaster} & Go, Gear & rilix & Free & device id & 21,635 & -- & 2 \\
        \texttt{Shuttle Commander} & Quest, SteamVR & immersive vr education & \$9.99 & genetic, health, political, belief & 158	& -- & \textasciitilde 300 \\
        \texttt{Pinball Fx2 VR} & Go, Rift, Gear, Quest & zen studios & \$4.99 & email, biometric & 358 & -- & \textasciitilde 200 \\
        \texttt{Storyup} & Go, Gear & storyup vr & Free & biometric & 188 & -- & \textasciitilde 250 \\
        \texttt{Dinosaurs} & Go, Gear & simdesign vr studio & \$0.99 & environment, eye tracking & 81 & -- & 394 \\
        \texttt{Stormland} & Rift & insomniac games & \$39.99 & mac, email, vr avatar & 1,934 & -- & 22 \\
        \texttt{Snapshot VR} & Sidequest, SteamVR & giant scam industries & Paid & password, user id, ip & 12.9k & 4.81 & \textasciitilde 300 \\
        \texttt{Realities} & Viveport & realitiesio & Free & user id, usage time, geo-location & 24 & 3.9 & \textasciitilde 100 \\
        \texttt{Easy Map 3d} & Microsoft & ivan fuligni & Free & photo, network information & 12 & -- & 10 \\
        \texttt{Bait} & Go, Viveport, Quest & -- & Free & device id, user id, gameplay & 24,364 & -- & 1 \\
        ... & ...  & ...            & ...  & ...   & ...  & ...\\
    \bottomrule
    \end{tabular}
    \label{tab:C4-violate}
\end{table*}

\end{document}